\documentclass{aa}

\usepackage{graphicx}
\graphicspath{{plots/}}
\newcommand{\be}{\begin{equation}}
\newcommand{\ee}{\end{equation}}

\newcommand{\kms}{$\rm {km}~\rm s^{-1}$}
\newcommand{\kmskpc}{$\rm {km}~\rm s^{-1}~\rm kpc^{-1}$}
\begin{document}

%\thesaurus{08(09.10.1; 09.13.2; 08.16.5)}
\title{Application of the global modal approach to spiral galaxies}
\author{V. Korchagin \inst{1,2}, N. Orlova \inst{2}, N. Kikuchi \inst{3},
       S. M. Miyama \inst{4}
    and A. Moiseev \inst{5}}
  \institute{
  Yale University, New Haven, Connecticut 06520-8101, USA\\
    \email{vik@astro.yale.edu} \and
  Institute of Physics, Stachki 194, Rostov-on-Don, Russia\\
    \email{nata@ip.rsu.ru} \and
  Earth Observation Research Center, National Space Development
Agency of Japan, 1-8-10 Harurau, Chuo-ku, Tokyo 104-6023, Japan\\
    \email{kikuchi@eorc.nasda.go.jp} \and
National Astronomical Observatory, Mitaka, Tokyo 181--8588, Japan\\
\email{miyama@th.nao.ac.jp} \and
Special Astrophysical Observatory of Russian AS, Nizhnij
       Arkhyz 369167, Russia \\
\email{moisav@sao.ru}
}
\offprints{V. \ Korchagin}
\date{received; accepted}

\abstract{We have tested the applicability of the global modal
approach in the density wave theory of spiral structure
for a sample of six spiral galaxies: NGC 488, NGC 628,
NGC 1566, NGC 2985, NGC 3938 and NGC 6503.
The galaxies demonstrate a variety of spiral patterns from
the regular open and tightly wound spiral patterns to a
multi-armed spiral structure.
Using the observed radial distributions of the stellar
velocity dispersions and the rotation curves we have
constructed equilibrium models for the galactic disks
in each galaxy and analyzed the dynamics of the spiral perturbations
using linear global modal analysis and  nonlinear hydrodynamical  simulations.
The theory reproduces qualitatively the observed properties of the spiral arms
in the galactic disks. Namely the theory predicts observed grand-design
spiral structure in the  galaxy NGC 1566, the tightly-wound spirals
in galaxies NGC 488 and NGC 2985, the two-armed spiral pattern
with the third spiral arm in the galaxy NGC 628, and the multi-armed spiral structure in
the galaxies NGC 3938 and NGC 6503. In general, more massive disks are
dominated by two-armed spiral modes. Disks with  lower mass, and
with lower velocity dispersion are simultaneously unstable for
spiral modes with different numbers of arms,
which results in a more complicated pattern.
\keywords{Galaxies: kinematics and dynamics -- Galaxies: spiral --
Interstellar Medium: structure --
Physical Data and Processes: Instabilities}
}

\titlerunning{Global modal approach}
\authorrunning{Korchagin et al.}

\maketitle

%%%%%%INTRODUCTION%%%%%%%%%%

\section{Introduction}
\label{introduction}

During the last 60 years many efforts have been made to reveal the nature
of the spiral structure in galaxies.
Most researchers agree that the spiral structure
is a manifestation of density variations
in the galactic disks under the influence of gravity.
However, opinions as to the governing mechanisms of
spiral structure differ beyond this starting point.
There is no consensus as to whether the spirals are a long-lived phenomenon, or are
regenerated many times during the galactic evolution.
C.C. Lin and his collaborators advocate long-lived
spiral patterns which are the manifestation of unstable
global modes (see, e.g., Bertin et al. 1989a,b).
Other researchers favor short-lived or recurrent patterns
which are developed through swing amplification in shearing flows
or by external forcing
(Goldreich \& Lynden-Bell 1965; Julian \& Toomre 1966; Toomre 1981).
A new realization of the recurrent mechanism was recently
proposed by Sellwood (2000) with further development
by Fuchs (2001a,b).

Some studies
demonstrate that the modal approach, and the fluid dynamical approximation
qualitatively
explain the dynamics of the collisionless galactic disks.
Vauterin and Dejonghe  (1996) performed a linear global modal analysis
of a family of collisionless self-gravitating disk models.
Kikuchi et al. (1997) using fluid approximation made a comparison of the linear global modal
analysis with the results of
Vauterin and Dejonghe (1996).
Kikuchi et al. (1997) showed that the
stability properties of disks described in a fluid approximation
are in good qualitative and to some extent quantitative
agreement with the properties of the global modes
found by Vauterin and Dejonghe from the exact collisionless Boltzmann equation.
However, only a direct comparison of the theoretical models
with the observed properties of the spiral patterns of
particular galaxies can help to make a choice in favor
of one or another theory.
A number of authors have undertaken such comparisons in the past
(e.g., Lin et al. 1969;
Roberts et al. 1975; Mishurov et al. 1979;
Elmegreen \& Elmegreen 1990).
Typically, these studies
relied on empirical estimates of the positions of the
corotation and Lindblad resonances in the galactic disks
using ``optical tracers'' in the optical images of the galaxies.
A local dispersion relation was used then to calculate the spiral
response under an additional assumption regarding the radial behavior of the
stability parameter $Q$ derived by Toomre (1964).
Arbitrary assumptions made in such comparisons
obviously reduce the predictive power of the theory.

Based on the recent long-slit spectroscopic observations of the disk galaxies,
we use a new approach to
predict theoretically the parameters of spiral structure in the galactic disks.
The approach
we use is based on the observational data.
Measurements of the radial distribution of the stellar velocity dispersion
in the galactic disks in combination with the known rotation curves provide the
disk's basic axisymmetric properties.
These data uniquely determine the
axisymmetric background equilibrium of the galactic disks, and can be used
for the linear-, and the nonlinear analyses aimed at modeling the spiral structure.

Such an approach have been applied to model spiral structure in the galaxy
NGC 1566 (Korchagin et al. 2000). The two armed spiral pattern
constitutes the most unstable global mode in the disk of NGC 1566, and the theoretical
surface brightness distribution and the velocity variations across the spiral arms
of NGC 1566 are in qualitative agreement with observations for the obsrtvationally
based models of NGC 1566.

In this paper we extend the study to a set of
nearby spiral galaxies of different morphological types.
Namely, we choose six spiral galaxies with
 measured kinematical properties of their disks: NGC 1566, NGC 488, NGC 628,
NGC 2985, NGC 3938 and NGC 6503.
We include the galaxy NGC 1566 studied previously by Korchagin et al. (2000)
in our list, for which we present some new results.
We use an approach similar to the one previously used to model the spiral structure
in the grand design spiral galaxy NGC 1566 (Korchagin et al. 2000).
We construct observationally-based axisymmetric
background distributions in the disks of the spiral galaxies
to undertake linear global modal analysis, and determine
the set of unstable global modes that might grow in their disks.
The results of the linear global modal
analysis are then compared with the direct nonlinear simulations of
the galactic disks using 2-D one-component
and multi-component fluid dynamical codes.
We follow the dynamics of the disks of the galaxies
starting from the random initial perturbations up through the
nonlinear saturated phase of spiral instability.
We compare then the theoretically predicted spiral patterns with the
observed properties of the spiral galaxies, and
find that the theoretically predicted global modes reproduce qualitatively well the observed
morphological properties of the spiral spiral structure in our sample.
We conclude that our results support the global modal approach as
a theoretical explanation of the spiral structure of galactic disks.

\section{Approach and Assumptions}

{\it Fluid approximation.}--We base our study on a fluid dynamical
approximation.  In this approach, a stellar disk is modeled as a fluid with the polytropic
equation of state:
\be
    P_s = K_s\sigma(r)^{\gamma}
\ee
Here $\sigma(r)$ is a surface density, $P_s$ is a vertically integrated pressure, $\gamma$ is the
polytropic index, and $K_s$ is a polytropic constant.

There are a few justifying arguments why the fluid dynamical equations
can be applied to describe the behavior of galactic disks
built mostly from collisionless stars. Marochnik (1966),
Hunter (1979) and Sygnet et al. (1987) derived from the
collisionless Boltzmann equation a set of hydrodynamic
equations that describe the behavior of perturbations in the
collisionless rotating disks.
Another argument comes from the
comparison of the global modes
calculated in a fluid-dynamical approach with the
results of the exact simulations that use the collisionless Boltzmann
equation. Kikuchi et al. (1997) found that the global
modes calculated in a collisionless disk by
using the Boltzmann equation (Vauterin and Dejonghe 1996),
are in a good agreement with the global
modes calculated in a fluid dynamical approach.

We use in this paper the simplest realization of
the collisionless hydrodynamics - the polytropic
equation of state with the polytropic index $\gamma =$ 2
(Marochnik 1966). The disks of galaxies are self-gravitating
in a perpendicular direction, and this law naturally
explains the empirical ''square root'' proportionality between
the velocity dispersion and the surface
density found in galactic disks (Bottema 1992).

{\it Velocity dispersion.}--Observational data
for the line-of-sight stellar velocity dispersion as a function of
a galactocentric radius are available for all galaxies in our sample.
The observed line-of-sight velocity dispersion
is determined by the radial,
tangential and vertical components of the velocity dispersion.
For the galaxies that have a low inclination,
the observed velocity dispersion gives approximately
the z-component of the velocity dispersion.
For the highly inclined galaxy NGC 6503,
the observed line-of-sight velocity dispersion is determined by the
tangential component of the velocity dispersion ellipsoid (Bottema 1989).

The observed radial distributions of the velocity dispersion
in the galactic disks are close to exponential profiles.
We assume that the radial distributions of the velocity dispersion
in the galactic disks are exponential and choose, in agreement with observations
(Bottema 1992),
the radial scale length of the velocity dispersion
to be twice as large as the radial scale length of the surface
density distribution in the galactic disks.
To construct the equilibrium models of the galactic disks,
one needs to know the radial as well as the vertical components
of the stellar velocity dispersion.
Two galaxies in our sample, namely NGC 2985
and NGC 488 have measured velocity ellipsoids (Gerssen et al. 2000; Gerssen et al. 1997),
and we adopt measured
radial and vertical velocity dispersions for these galaxies in our analysis.
For the rest of the sample, we  accept the value of the ratio
of vertical to radial velocity dispersion of stars in the galactic disks,
and assume this ratio to be constant throughout a galactic disk.
In general, the ratio of the velocity dispersions $c_z / c_r$
varies with the Hubble type of the galaxy, and is in the range
$0.5 - 0.8$ (Merrifield 2001; Shapiro et al. 2003).
To construct equilibrium models for the galaxies with unknown velocity dispersion ellipsoids,
we vary the ratio of the vertical to the radial velocity dispersions
between $0.5 and 0.8$. We find, however,
that once the galactic disk is unstable, the morphological
properties of the growing spiral modes, i.e., the number of arms and their pitch angles,
are largely insensitive to the adopted velocity dispersion ratio.

{\it Surface density.}-- The surface density in the galactic disks
is determined under the assumption that the disks are
in equilibrium in the vertical direction, and can be modeled
as a locally isothermal, self-gravitating slab. The surface
density $\sigma(r)$ and the vertical velocity dispersion $c_z$
are then related as (Spitzer 1942):
\be
\sigma(r) = c_z^2/\pi G z_0
\label{eq2}
\ee
Here $G$ is the gravitational constant, and $z_0$ is the exponential
scale height of the disk density distribution.

The exponential scale height of the galactic
disks $z_0$ is an unknown parameter for the galaxies in our sample.
Observations of a few edge-on galaxies (van der Kruit \& Searle 1982),
show that the vertical scale height is constant with radius,
and varies between 0.3 - 0.8 kpc within their sample.
The measurement of the vertical scale height of the thin disk component
in the Milky Way gives a value close to 0.28 kpc (Drimmel and Spergel 2001).
We assume this parameter to be constant throughout the galactic disk,
and vary it in our models within 0.3 - 0.8 kpc.

In the outer galactic disks, the stars that are observed, are relatively bright and massive.
These stars do not contain most of the mass of an old stellar disk which is accumulated in
form of dim  K - M dwarfs. Nevetheless, the method correctly predicts the
local surface density of the disk. One can use the measured velocty
dispersion and the scale height of a `tracer' stellar population
to estimate the total surface density of underlying mass distribution ( Korchagin et al. 2003).

The
radial scale lengths of the exponential disk are
photometrically determined for each particular galaxy.
For numerical reasons, we assume that the surface density vanishes at the outer
boundary of the disk and introduce a dimensionless factor $[1-(r/R_{out})^2]^5$
which smoothly enforces this behavior.

The density distributions in the central regions of the galaxies NGC 488
and NGC 2985 are dominated by the bulge.
To study a possible influence of the bulge on the modal properties
of the disks in these galaxies, we followed Bertin et al.
(1989a), and introduce in some models another dimensionless multiplier which
decreases the surface density in the central regions
mimicing a replacement of the disk stars by a bulge stellar populations.
We find though  that a drop of the disk density near its center
has a little effect on the disk modal properties.

{\it Rotation curve.}--
To model the observed rotation curves in our sample of galaxies, we have adopted
the rotation curve from Korchagin et al. (2000):
\be
v_0(r) = {V_1 r \over (r^2 + R_1^2)^{\alpha}}
       + {V_2 r \over (r^2 + R_2^2)^{\beta}}
\label{eq4}
\ee
where the constants $V_1, V_2, R_1$ and $R_2$ are determined from the best fit to the
observed rotation curve in a particular galaxy, and the constants $\alpha$ and $\beta$
in most of cases were equal to 0.75.
Analytical representation to the rotation curves of the galaxies
exhibits basic features of the galactic rotation,
and allows an accurate modelling of the observed rotation of the galactic disks.
We find that a particular choice to represent analytically the rotation
curve of a galaxy is not important once a modeled rotation curve is close
to an observed rotation profile of the galactic disk.

{\it Q-parameter.}-- Radial velocity dispersion together with
the surface density and epicyclic frequency
give the radial dependence of the stability Q-parameter
(Toomre 1964), given by:
\be
Q = {c_r(r) \kappa(r) \over 3.36 G \sigma(r)}
\label{eq5}
\ee

Here $\kappa(r)= \sqrt { 2 {V \over r} ( {V \over r} + {d V \over d r} ) }$ is the epicyclic frequency
determined by the observed rotation curve, $ \sigma(r)$ is the disk surface density,
and $c_r$ is the radial velocity
dispersion in the galactic disk. With
exponential distributions of the surface density and the velocity
dispersion, the Q-parameter tends to be larger in the central regions and at the
peripheries of the galactic disks. The minimum value of the Q-parameter is larger than unity
in the galactic disks which, however, does not prevent them from experiencing global instabilities.

{\it Inactive halo.}--
An equilibrium rotation of the galactic disk is supported by the combined disk and halo/bulge gravitational
potential. Once the radial profile of surface density of a galactic disk is determined from
the observed velocity dispersion, the gravitational potential and the density distribution
of an inactive spherically symmetric halo and bulge can be calculated from the rotational
equilibrium of the disk.  The potentials of spherical components are kept intact during
the all simulational run. Typically, the mass of spherical components in computational
region is a few times larger than the mass of a galactic disk.

{\it Cold gas component.}--
We explore the simplest possible model of the galactic disks approximating
them as a one-component fluid obeying an adiabatic equation of state.
Galactic disks consist of stars and gas in different forms, that
affects the stability properties and the dynamics of the unstable modes.
This question was addressed by a number of studies (e.g., Jog \& Solomon 1984,
Berting \& Romeo 1988, Orlova et al. 2002). An admixture of a cold gas component
increases the growth rates of the unstable global spiral modes in the disk.
However, the appearance of the global modes in the multi-component models
remains quite similar to that in one-component models.
E.g., an unstable two-armed spiral growing in a one-component disk
remains a governing unstable mode in a multi-component disk, and has
a comparable winding (Orlova et al. 2002).
The colder components would not change therefore a qualitative analysis
we present.

{\it Computational methods.}--
The computational methods used in this paper are similar to those described in
Korchagin et al. (2000). Briefly, the linearized hydrodynamical equations
are reduced to a single integro-differential equation that describes an unstable global mode
( e.g., Adams et al. 1989). Together with the boundary conditions,
this equation formulates the eigenvalue problem, which is solved numerically
by means of a matrix method. In this method, the governing eigenvalue equation
is applied at (N+1) radial grid points, and the problem is then reduced
to an (N+1)$\times$(N+1) matrix equation which yields an eigenfrequency of the
global mode.
The nonlinear simulations are made with the two-dimensional numerical code based
on a second-order van Leer-type advection scheme to integrate hydrodynamical equations,
and a fast Fourier transform algorithm to solve the gravitational potential.
For the runs discussed here, a grid of 256 $\times$ 256 zones was employed
with equally spaced azimuthal zones, and logarithmically spaced radial zones.
We checked a dependence of the results on grid resolution by employing the grids with resolutions
64 $\times$ 64, 128 $\times$ 128 and 512 $\times$ 512.
We find that results of the simulations do not change if the grid resolution
128 $\times$ 128 or higher.

The simulations were launched from an initial condition in which each hydrodynamic
zone was given a random density perturbation of up to one part in $10^6$ away
from the equilibrium state. In some cases, random density perturbations
of order of $10^{-15}$ of unperturbed state were imposed.
The growth rate, patterm speed, winding
or the nolinear saturation amplitude of an emerging spiral pattern do not depend
on amplitude of the initial noise perturbations.

%%%%%%%%%%%%%%%%%%%%%%%%Results%%%%%%%%%%%%
\section{Results}

\subsection{NGC 1566}

With its near-perfect two-armed spiral pattern and with inclination
of about $26\degr$ , this nearby ( distance 17.4 Mpc) spiral galaxy was studied observationally
by a number of authors. Korchagin et al (2000) have undertaken
a theoretical modeling of spiral structure in NGC 1566
based on the Bottema's (1992) measurement of the rotation curve
and the radial dependence of the $z$-component of the velocity
dispersion in its disk.
Assuming a fixed vertical scale height of 0.7 kpc, Korchagin et al. (2000)
found that NGC 1566 should be gravitationally unstable
when the ratio
of the vertical to the radial velocity dispersions, $ c_z/c_r$ is close to unity.
In this paper we make different assumptions for NGC 1566, choosing a lower
values of $ c_z/c_r$, and varying a vertical scale height of the disk.
Measurements of the Milky Way velocity ellipsoid in solar neighborhood
give for this ratio the value of about 0.5 ( Dehnen and Binney, 1998).
Theoretical arguments (e.g., Villumsen 1985; Lacey 1984) also
favor a smaller value
of 0.6 -- 0.8.
There is an observational evidence for a systematic trend of $c_z/c_r$ ratio
among galaxies of different Hubble types (Merrifield 2001; Shapiro et al. 2003). Galaxies of late
Hubble types tend to have a lower $c_z/c_r$ ratio, while in the galaxies of earlier
Hubble types this ratio is larger.  Shapiro et al. (2003) studied the ratio of the
vertical to radial velocity dispersion for four spiral galaxies of Hubble type from Sa to Sbc.
They find that $c_z/c_r$ is generally in the range of  $0.5 - 0.8$.
In this paper, we re-analyze the theoretical models
of spiral structure in NGC 1566 varying the value of the vertical to radial
velocity dispersion within $0.5 - 0.8$,
but allowing the vertical scale height of the disk of NGC 1566
to vary within $0.3 - 0.75$ kpc.

\begin{figure}
   \resizebox{\hsize}{!}{
     \includegraphics[angle=-90]{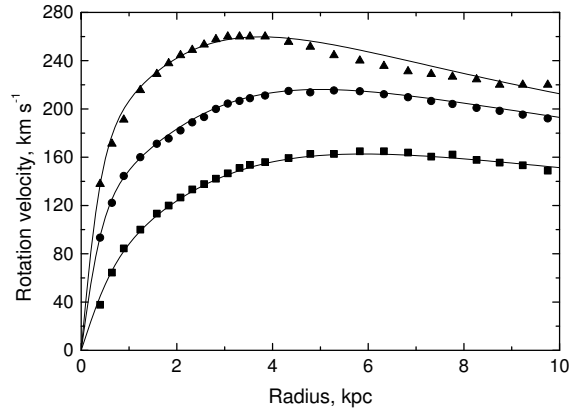}
   }
   \caption{The rotation curve of NGC 1566 as modeled by the equation 3.
The curves are superimposed on observational
data taken from Bottema (1992). The ``upper'' and ``lower'' curves arise from the $\pm$ 5
deg uncernainty in the deremining of inclination of NGC 1566. Parameters
of the rotation curves are given in Table 1 of  Korchagin et al. (2000).
    }
   \label{fig01}
\end{figure}

The observed velocity dispersion in NGC 1566 is approximately four percent larger than
the $z$-component of the velocity dispersion (Bottema 1992). Following Bottema,
we neglect this difference and assume that measured velocity dispersion  gives
the real $z$-dispersion in the disk of NGC 1566.

\begin{figure}
   \resizebox{\hsize}{!}{
     \includegraphics[angle=-90]{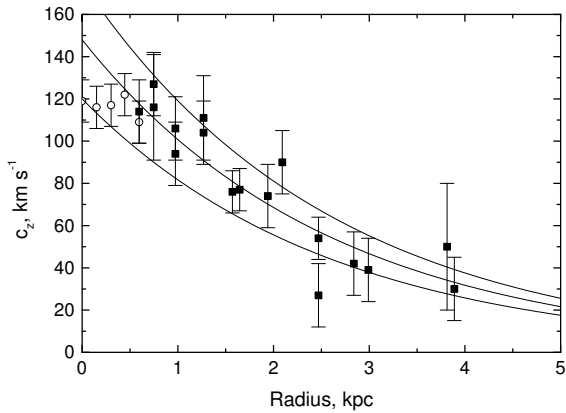}
   }
   \caption{The observed velocity dispersion of stars in the disk of NGC 1566
taken from Bottema (1992). Superimposed are exponential velocity
dispersion distributions with the radial scale length of 2.6 kpc,
and the values of the velocity dispersion at the disk center of 120, 148 and 176 km/sec.
    }
   \label{fig02}
\end{figure}

Figure 2 shows the observed
vertical velocity dispersion as a function of radius taken from Bottema (1992)
modeled  with an exponential distribution. The radial scale length for
the velocity dispersion profile was choosen of 2.6 kpc, which is twice as large than
the scale lengh of the surface brightness distribution (Bottema 1993).
The best fit to the exponential function gives the value for the velocity
dispersion at the disk center of 148 km/sec. An exponential function
satisfactory reproduces the observed profile
of the velocity dispersion within the error bars except for the near-central regions (Figure 2).
An anomalous behavior of the velocity dispersion at the core of NGC 1566
is probably related to a small central bulge (Bottema 1992).

We considered also models with the ''upper'' and ''lower'' velocity dispersion
profiles allowed within the error bars, as shown in Figure 2.
These models have central velocity dispersions of 120 km/sec and 176 km/sec.

Knowledge of the $z$-component
of the velocity dispersion allows us to determine the surface density in the disk,
provided the vertical scale height of the stellar distribution
is known. The vertical scale height of the stellar disk of NGC 1566, which is seen nearly face-on,
cannot be measured directly. Studies of the surface photometry of edge-on spiral galaxies
show that the vertical disk profile shape is independent of galaxy type, and varies
little with position along the major axis (van der Kruit and Searle 1982; de Grijs et al. 1997).
Van der Kruit and Searle (1982) find the average  vertical scale height
of the old stellar population in  galactic disks, when fitted to the sech$^2$- function, is about 0.7 kpc.
This value is higher than the sech$^2$ vertical scale height measured in the Milky Way galaxy.
Recent determinations of the thin disk scale height of the Milky Way in the solar neighborhood
give the value of about 280 pc (Drimmel and Spergel 2001).
We allow therefore in our models the vertical scale height to vary within 0.3 - 0.75 kpc.
We find that the morphological properties of the emerging spiral pattern
depend very little on the assumed value of the vertical scale height of a galactic disk, as
demonstrated in the following section.

\subsubsection {Linear Models}

Table 1 lists the parameters of the twelve models of NGC 1566 studied in our linear analysis.
 All the models have the fixed 'medium' rotation curve with the maximum rotation velocity
of 212 km/sec. The range of disk vertical scale height values is 0.3 - 0.75  kpc,
so the minimum value
of Toomre $Q$-parameter is about 1.4 - 1.6.
The last column of the Table 1 gives the pattern
speed, $\Omega_p$, and the growth rate, Im$ \omega$ of the most unstable $m=2$ spiral global mode.
The ratio of the growth rate to the pattern speed $\omega_p$ varies
from approximately one-fourth for the model with the $Q_{min} = 1.42$, to the one-tenth
for the model with the $Q_{min}$ equal to 1.6.
The shape of the spiral patterns does not change considerably
from model to model, but
the corotation radius and the overall extent of the spiral
patten changes.
Model $f$ from Table 1, which has intermediate values of key parameters,
is used in our nonlinear simulations.

{\it Dependence on the disk thickness.}-- The dependence of
the global modes on the vertical scale height of the disk was studied in the models
built from model $f$ by varying the
disk scale height. The disk scale height was varied
from 0.3 kpc, when the minimum value
of Toomre's $Q$-parameter is less than unity, to 0.5 kpc, until the
disk becomes totally stable.
The two-armed spiral remains the most unstable global mode in the sequence of models
built in such way.
We find that variation of the disk thickness within an observationally allowed range
does not affect the
modal properties of the disk. The winding
of spirals do not change considerably, however, the growth rate increases by a factor of six,
as the scale height is decreased.
By decreasing the vertical scale height of the disk we simultaneously
increase the disk mass from $3.5 \times 10^{10}$ M$_{\odot}$
to $5.1 \times 10^{10}$ M$_{\odot}$.

{\it Dependence on the rotation curve.}--
Together with the 'standard' medium rotation curve, we considered also models
with the ''upper'' and ''lower'' rotation curves
allowed within the error bars, as shown in Figure 1.
These models demostrate qualitatively similar behavior.
The models have the ratio of the velocity dispersions $c_z/c_r$, fixed at 0.6,
and central velocity dispersion fixed at 148 km/s.
The rotation curve is varied from the ``lower'' to the ``upper'' curve as shown in Figure 1.
Again, while the rotation speed and the growth rate of the most unstable spiral are sensitive
to the particular parameters of the models (the rotation curve), the two-armed spiral remains the
most unstable global mode, and its morphological properties
vary little from model to model.

\subsubsection {Nonlinear simulations}

This section presents the results of the 2-D nonlinear simulations of the spiral
pattern which develops in the disk of NGC 1566 once it is seeded with small amplitude noise perturbations.
We choose as an example the model $f$ for our nonlinear simulations which is
built for the best fit medium rotation curve and for the best fit to the observed radial
distribution of the velocity dispersion.

In accordance with the linear analysis,
the model is the most unstable with respect to the growth of a two-armed spiral.
The perturbations exponentially
grow during approximately $10^9$ years when the exponential growth phase merges into
a saturation phase. The overall dynamics of the perturbations is dictated by the
two-armed spiral mode.
We checked that by calculationg time dependence of the global perturbation amplitudes,
determined as the Fourier amplitudes of the perturbations averaged throughout the disk:
\be
A_m \equiv {1 \over {M_{d}}}
\left\vert \int_{0}^{2 \pi} \int_{R_{in}}^{R_{out}}
\sigma(r,\phi) r dr \, e^{-im\phi} d\phi \right\vert
\, .
\label{eq6}
\ee
Here $M_d$ is the mass of the disk, $R_{in}$ and $R_{out}$ are the radii of the inner and the outer
disk boundaries, and $m$ is the number of arms.
Figure 3, which shows the time sequence
for the surface density perturbations developing in the disk, also illustrates that
the two armed pattern dictates the dynamics of perturbations.
The nearest competitor, the the three-armed spiral, although present in the 2-D  density
perturbations, does not play a significant role in the morphological properties
of the developing perturbations.

 {\it Dependence on boundary conditions.}--
Despite a good quantitative agreement between the results of the linear
modal analysis and the nonlinear simulations,
the limited radial dynamic range and the reflective boundary conditions accepted in
simulations can affect our results.  It is known (Toomre 1981), that
a steep truncation of the density distribution at the outer edge of the disk can alter the spectrum of the global modes of the disk, thus compromising the comparison of
theoretical models with observations. Laughlin et al. (1998) studied
the behavior of perturbations in the nonlinear simulations
depending on boundary conditions as well as on the radial dynamic range of the models.
They find that the early development of a radially extended disk is more complex
than the  evolution of the model with the lower radial range, but the essential features of
the two disks are very similar, and the two-armed spiral is the dominant feature
in both experiments.

\begin{figure*}[tp]
  \centerline{\hbox{
  }}
  \caption{Two-dimensional contour plots for the density perturbations taken at time 0.66$\times$ 10$^9$
years in model $f$ with the varying outer boundary condition. Panel $a$ -
the outer boundary is at 20 kpc,  and the surface density is vanishing at the outer
boundary. Panel $b$ - the outer radius is at 20 kpc, but the surface
density is ubruptly truncated. Panel $c$ - model with an exponential density distribution,
but with an outer radius extended to 40 kpc.
   \label{fig03}
   }
\end{figure*}

Figure 3 confirms the result of Laughlin et al. (1998). The panel $a$) in Figure 3 shows
the two-dimensional contour plot taken at time $0.66 \times 10^9$ yrs for the perturbations
developing in model $f$. In this model, the
outer boundary is placed at at 20 kpc
with the surface density chosen to be vanishing at the outer boundary in
a smooth way . The panel $b$) of the same Figure shows the spiral pattern
developing in the model with the outer radius placed at 20 kpc, but with the
exponential density distribution abruptly
truncated at the outer edge of the disk.
The panel $c$) in  Figure 3 is a result of simulations performed with the
disk which has an exponential density distribution inside $R_{out} = 20$ kpc,
but with the outer boundary moved out to $R_{out} = 40$ kpc.
The surface density in the region between $r=20$ kpc and $r=40$ kpc
is set to a small value about $10^{-6}$ of the central surface density of the disk.
In all three frames the perturbations
are shown at the same moment of time. As it can be seen from Figure 3,
the two-armed spirals are
nearly identical in all three cases, and the outer boundary condition as well as the
radial dynamic range of the disk do not affect the behavior of perturbations.

\subsubsection{Comparison with observations}

Figure 4 shows an I-band image of NGC 1566 taken from Korchagin et al. (2000) (left frame),
as compared to the perturbed density distribution developing in model $f$ at a
time of $0.66 \times 10^9$ years.
The spiral pattern found in the linear global modal
analysis, and the pattern emerging in the disk  from
the noise perturbations agree well with each other,
and with the observed spiral in
NGC 1566. The pitch angle of the theoretical pattern is about
35$\degr$, which is in  good agreement with the pitch
angle of  36$\degr$ measured by Vera-Villamizar et al.(2001) in NGC 1566.
Our simulations show that the two-armed spiral
dominates the dynamics of the perturbations during the linear,
as well as the nonlinear saturated stage of pattern evolution.
The time of the nonlinear phase is comparable to the time of the exponential
growth phase. In total, the modeled two-armed pattern exists more than 10$^
9$ years, and makes about eight revolutions during the simulation time.

\begin{figure*}[tp]
  \centerline{\hbox{
  \includegraphics[width=\textwidth]{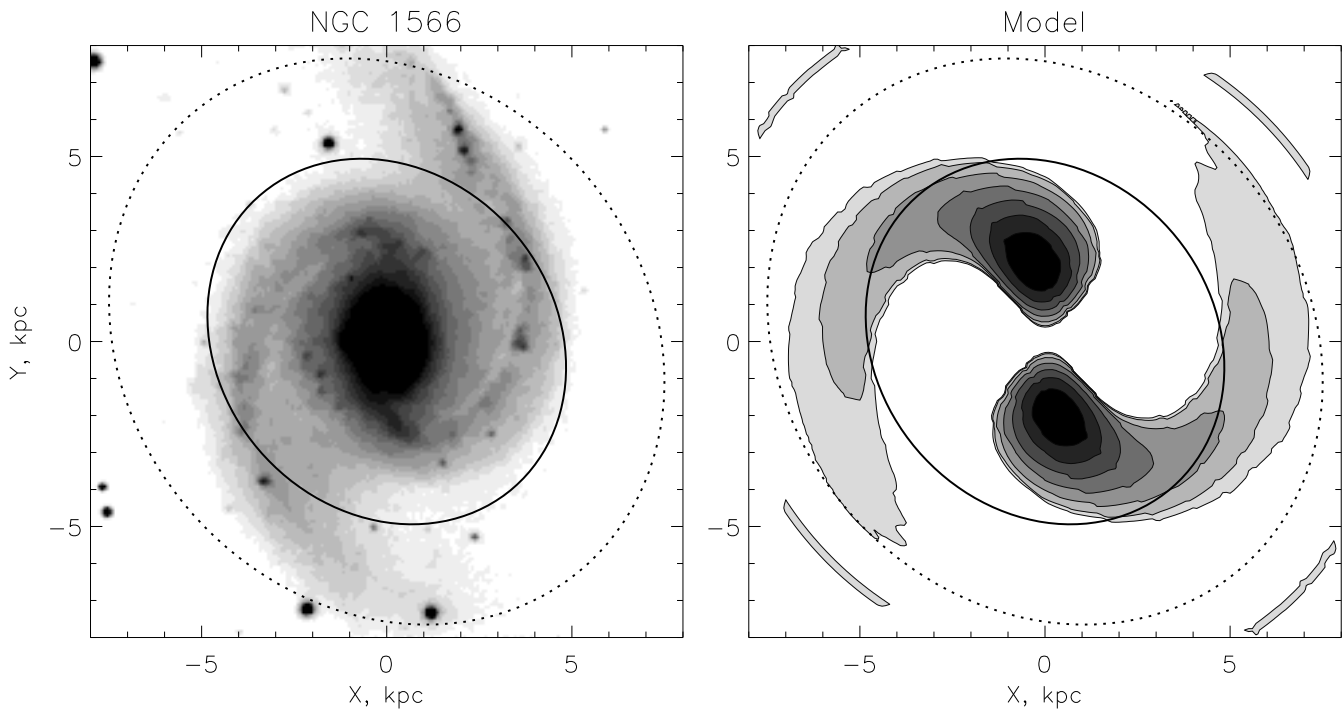}
  }}
  \caption{Left panel: an I-band image of NGC 1566 taken from Korchagin et al. (2000). Right panel:
the theoretical density distribution taken from the nonlinear simulation of model $f$ at
time 0.66$\times 10^9$ years. The contour levels
are logarithmically spaced between the maximum
value of the perturbed density and 1/100 of
the maximum perturbed density.
The solid line shows the position of the corotation resonance,
and the dashed line shows the
position of the outer Lindblad resonance.
   \label{fig04}
   }
\end{figure*}

{\it Corotation resonance.}-- The linear, as well as the nonlinear simulations
provide a value of the pattern speed, which in principle
allows the determinatin of the positions of the
corotation and the Lindblad resonances.
We measure in our simulations of NGC 1566 a pattern speed of
33 \kmskpc,
which places the corotation resonance at approximately 6 kpc.
Based on optical features of NGC 1566 (Elmegreen \& Elmegreen 1990) and by
comparing of blue and infrared frame azimuthal profile Fourier
transforms of disk surface brightness (Vera-Villamizar et al. 2001)
the position of the corotation resonance in NGC 1566 was determined at 8.5 - 10 kpc.
There are a few possible sources for such a discrepancy.
In our modeling of NGC 1566, we used the rotation curve
taken from Bottema (1992). Bottema's measurements give a maximum value
of the rotation speed of 212 $\pm$ 30 km s$^{-1}$.
Vera-Villamizar et al.(2001) based their studies on the rotation curve
taken from Persic $\&$ Salucci (1995) which has a maximum
rotation speed of 150 \kms. The uncertainties
in the measurement of the rotation curve obviously alter the theoretical
prediction of the positions of the principal resonances in the disk of NGC 1566.

As was demonstrated above, the overall morphological properties of the most
unstable global mode do not
depend strongly on the minimum value of Toomre's Q-parameter once
the galactic disk becomes unstable towards the growth of global modes.
The growth rate, and the pattern
speed do, however, depend on the minimum value of the Q-parameter.
We find that  more slowly growing spirals have lower pattern speed compared to
the fast-growing spirals. E.g., a pattern speed of 33 \kmskpc
 was found in the disk with a Q$_{min}$
of 1.6, while a disk with a Q$_{min}$ of 1.17 has a
pattern speed of 56 \kmskpc, and a corotation
resonance at 4 kpc. Therefore the uncertainties in the
measurements of the background properties of the disks prevent accurate
quantitative predictions of the positions of principal resonances.

\subsection{NGC 488}

NGC 488 is a nearby Sb galaxy with an inclination of about
40$\degr$ (Gerssen et al. 1997). With Hubble
constant of 75 km s$^{-1}$ Mpc$^{-1}$,
the distance to NGC 488 is about 30 Mpc.
Sandage \& Bedke (1994)
describe this galaxy as a prototype of the multi-armed spiral galaxies.
Hubble Space Telescope observations (Carollo et al. 1997, Sil'chenko 1999) reveal, however,
a tightly wound two-armed spiral pattern shown on the left panel of Figure 5.
This panel reproduces HST image of this galaxy in the optical band (F606W filter)
taken from the NED archive. To trace better the spiral structure in the central regions of the disk,
we have subtracted the bulge component assuming
a Sersic (1968) profile for the bulge,  $I(r)=I_0exp(-(r/r_0)^{1/n})$ with $n=2$,
and an exponential surface brightness distribution for the disk of NGC 488.

\begin{figure*}[tp]
  \centerline{\hbox{
  \includegraphics[width=\textwidth]{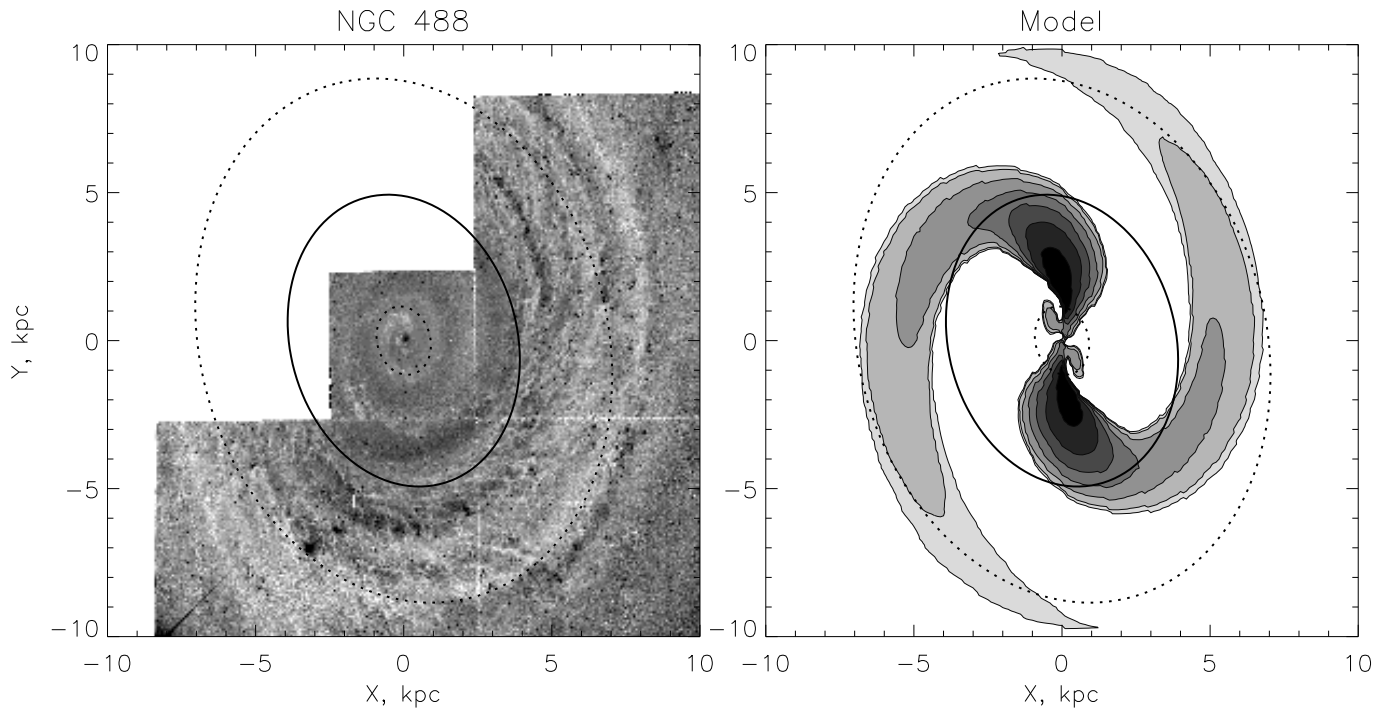}
  }}
  \caption{The left panel shows an HST optical band image (F606W filter)
of the galaxy NGC 488, taken from NED.
The image was decomposed into bulge and disk components, and
the bulge contribution was subtracted. The right panel shows the
theoretical density distribution in the disk of NGC 488 taken at time
$t=4\times 10^8$ yr. The contour levels
are logarithmically spaced between the maximum
value of the perturbed density and 1/100 of
the maximum perturbed density.
The solid line shows the position of the corotation resonance,
and the dashed line shows the
position of the outer Lindblad resonance.
   \label{fig05}
   }
\end{figure*}

NGC 488 is well studied kinematically.
Peterson (1980) measured the rotation curve of this galaxy using
emission line spectra and Gerssen et al. (1997) have measured
the disk rotation using stellar absorption-line data.
NGC 488 has a remarkably high rotation velocity with a maximum value
of about 360 \kms. Figure 6 shows both rotation curves
together with the best fits to the measurements.
In the outer parts of the galactic disk the stellar rotation curve is
consistent with the emission line measurements by Peterson (1980).
In the bulge region, gaseous emission line measurements give higher values for the disk
rotation. A discrepancy of gaseous and stellar rotation curves in the inner disk regions
results probably due to a large asymmetric drift of high velocity dispersion stars
in the bulge region. We explored both rotation curves to model stability
properties of the disk of NGC 488.

Gerssen et al. (1997) measured the components of the stellar velocity
dispersion. They found exponential radial distributions
for the radial and vertical components of velocity dispersion,
and using stellar absorption-line data estimated
the radial and the vertical velocity dispersions of stars at the
galactic center to be 253 $\pm$ 32 \kms and
164 $\pm$ 27 \kms respectively giving the value
of the ratio of vertical to radial velocity dispersion in the disk
of 0.70 $\pm$ 0.19.

\subsubsection{Linear Models}

Based on the measured stellar velocity dispersions, we have studied stability properties
for the family of models of NGC 488. The models listed in Table 2, explore the rotation curve
based on the emission line measurements of Peterson (1980).
The models cover an observed range of the
radial velocity dispersions at the disk center of 221 - 285 km/s.
The surface densities at the center of the disk
determined with the help of Equation (2), correspond to the observed range of the
disk vertical velocity dispersions of 137 - 191 km/s and the vertical scale  height
of the disk density distribution of 0.35 - 1 kpc.

The models are stable
when the minimum value of Toomre's parameter exceeds approximately 1.7.
All the unstable models demonstrate, however, quite similar behavior:
they are are mostly unstable towards a two-armed spiral.

\subsubsection{ Nonlinear Simulations}

{\it Exponential disk.}-
As an example, we present here the results of our nonlinear
simulations for the model $p$ that has an exponential surface density
distribution.
The total mass of the disk is 1.25 $\times$ 10$^{11}$ M$_{\odot}$.
The value of the vertical velocity dispersion at the disk center is 164 km/s.
The ratio of the radial to vertical velocity dispersion  is 0.73
as determined by Gerssen et al. (1997). Following Gerssen et al. (1997),
we assume that the disk velocity dispersion exponentially decreases
with radius with a radial scale length of 5.5 kpc.

Gerssen et al. (1997)
find that the kinematic scale length in the disk of NGC 488 is comparable to the
B-band photometric scale length, which is not the expectation
for the local isothermal approximation. They note however,
that a likely explanation is the fact that
the stellar mass distribution is better traced in the near-infrared K-band.
Empirically it is known that the scale lengths in K-band are
shorter than in B-band up to a factor of 2 (Peletier et al. 1996),
and we accept therefore the radial scale length for the surface density
distribution being two times shorter than the kinematic
scale length of Gerssen et al. (1997).

The equilibrium disk was initially seeded with the
noise density perturbations with amplitude of 10$^{-6}$ of the unperturbed density.
The exponentially growing
two-armed spiral pattern emerges from the noise perturbations.
Until the nonlinear saturation phase, the dynamics of perturbations
is governed by the two armed spiral.
During the nonlinear saturation phase, the spiral shocks develop,
and the regular spiral pattern is disrupted.
The hydrodynamical approximation is not adequate
to describe the behavior of the collisionless stellar disks
at highly nonlinear stages of evolution.

{\it Stellar rotation curve.}--
As it was mentioned above, there is a considerable discrepancy of gaseous and stellar
rotation curves in the inner disk regions.
To study how the results depend on our choice of the rotation curve, we used
stellar rotation curve to model the disk dynamics.
We find, that the difference between the both rotation curve measurements
do not affect an overall dynamics of the disk. Figure 7 shows a time sequence
for the density perturbations developing in model $p$, but with the
'stellar' rotation curve based on measuremenmts by Gerssen et al. (1997).
A two-armed spiral quickly emerges from the noise and governs the dynamics of perturbations
up to a nonlinear saturation phase.

\begin{figure}
   \resizebox{\hsize}{!}{
     \includegraphics[angle=-90]{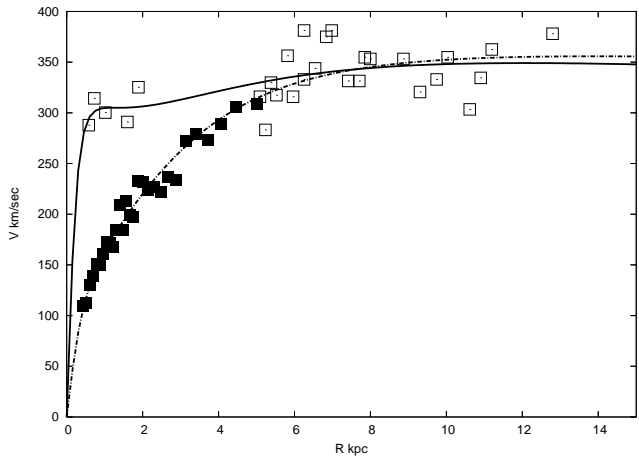}
   }
   \caption{Solid line - the best fit to the rotation curve of NGC 488 based on measurements by Peterson (1980)
(open squares) as modeled by the equation (3). The dimensionless parameters of the rotation curve
are $\alpha =$ 0.53, $\beta =$0.61, $V_1 =$1.59, $R_1 =$3.5, $R_2 =$0.17.
Dashed line - the best fit to the stellar rotation curve measurements
obtained by Gerssen et al. (1997) (filled squares). The dimensionless parameters of
stellar rotation curve: $\alpha =$0.53, $\beta =$ 0.61,
$V_1=$ 2.38, $V_2=$0.61, $R_1=$ 2.32, $R_2=$ 0.25. Velocity unit - 148.5 km/s, distance unit - 2 kpc.
    }
   \label{fig06}
\end{figure}

{\it Bulge.}--
NGC 488 is classified as an Sab galaxy and has a noticeable bulge.
A bulge-disk decomposition for NGC 488 by Kent (1985) gives
the value for the half-mass radius of the bulge of about 4.3 kpc
with the Hubble constant of 75 km/s/kpc.
Fuchs (1997) finds somewhat smaller value for the half-mass radius of about 2.5 kpc.

Little is known about the structure and the very existence of the disk inside the bulge.
As it was mentioned by Bertin et al. (1989a), the disk might be actually
replaced by the bulge stars in the inner regions, and the disk density
is reduced to insignificant values near its center.
We studied such a possibility by modeling the dynamics of a
disk that has a suppressed density distribution near its center.
We used the bulge density distribution from
Fuchs (1997), namely
\be
  \rho_b = \rho_{b0}\Big(1 + r^2/r_c^2 \Big)^{-1.75} {\mbox ,}
  \label{eq7}
\ee
with the bulge core radius $r_c$ of 0.92 kpc.
The resulting disk surface density  is shown in Figure 8.

\begin{figure}
   \resizebox{\hsize}{!}{
     \includegraphics[angle=0]{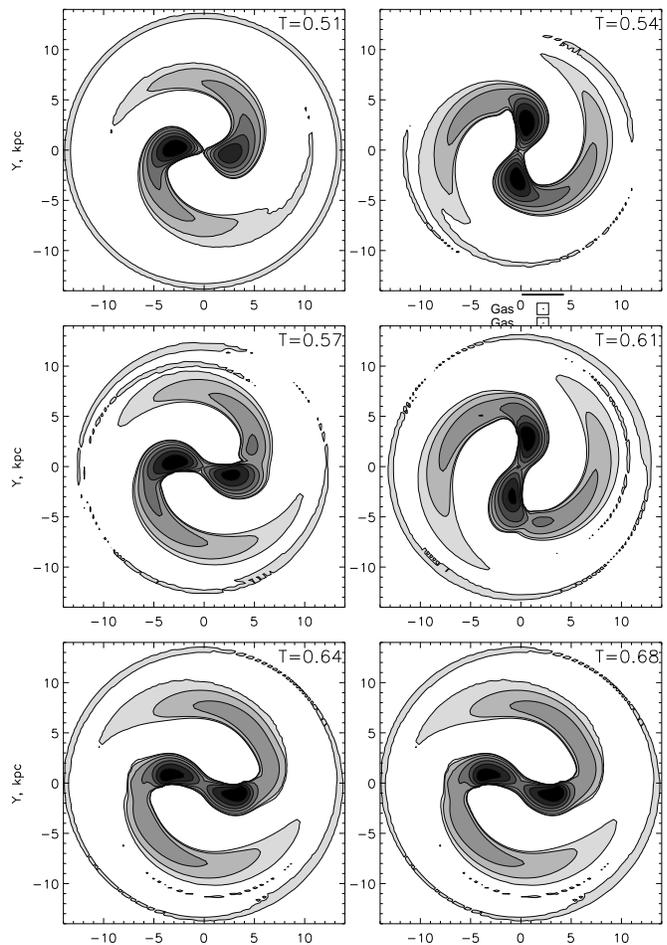}
   }
   \caption{The time sequence for the density perturbations developing in model $p$ with the 'stellar'
rotation curve based on measurements by Gerssen et al. (1997) shown in Figure 6.
    }
   \label{fig07}
\end{figure}

The disk dynamics is
similar to the two
previous cases. The disk is unstable towards the two-armed
spiral, and the spiral morphology does not differ much
compared to the two previous cases.
We simulated the dynamics of the disk using the gaseous as well as the stellar
rotation rotation curve, using the bulge parameters
taken from Kent (1985).
In all cases, the dynamics of perturbations
is governed by the two-armed
spiral. We find, however, that the models with the
density depression near the disk center exhibit more tightly wound spirals.

\begin{figure}
   \resizebox{\hsize}{!}{
     \includegraphics[angle=-90]{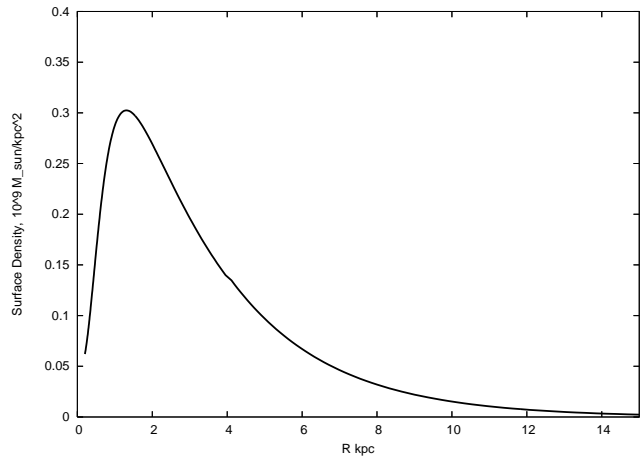}
   }
   \caption{The disk density distrubution in the model of NGC 488 with the disk density
suppression because of the central bulge.
    }
   \label{fig08}
\end{figure}

{\it Comparison with observations.}--
The right panel of Figure 5 shows the result of our nonlinear simulations
for model $p$ taken at time  4$\times 10^8$ yr. The Figure shows the contour plots
for density perturbations in the disk with the contour levels
logarithmically spaced between the maximum value
of the perturbed density and 1/100 of the maximum value of  perturbations.
As in the real galaxy,
the model produces a two-armed spiral pattern.
The modelled spiral arms are more open compared to the observed ones.
A discrepancy between the modeled
spiral arms and the observations, especially in the central disk regions,
is not surprising taking
into account the uncertainties in the observed density and the velocity  dispersion
distributions. However, within
all possible uncertainties of the observed equilibrium disk distributions, and
despite of our ignorance of the disk properties close to the center, all the models
support the two-armed spiral that governs the dynamics of perturbations for about one Gyr.
During this time, the spiral pattern makes about seven revolutions.

%%%%%%%%%%%%%%%%%%%%%NGC 628%%%%%%%%%%%%%%%%%%%%%%%%%%%%%%%%%%%%%%%%%%%%%%%%

\subsection{NGC 628}

The proximity (distance of about 10 Mpc) and the orientation of NGC 628
makes it a convenient object to study the spiral structure.
NGC 628 is a grand design galaxy that has two principal spiral arms,
and a number of secondary arms.
The I-band image taken from Larsen \& Richtler (1999) and shown in the left panel of Figure 9.

Fourier analysis of the spiral structure of NGC 628 by Puerari \& Dottori (1992)
shows that the two-armed spiral pattern is a dominating component.
However, there is a considerable admixture of a three-armed spiral in the
inner regions of the galaxy.

\begin{figure*}[tp]
  \centerline{\hbox{
  \includegraphics[width=\textwidth]{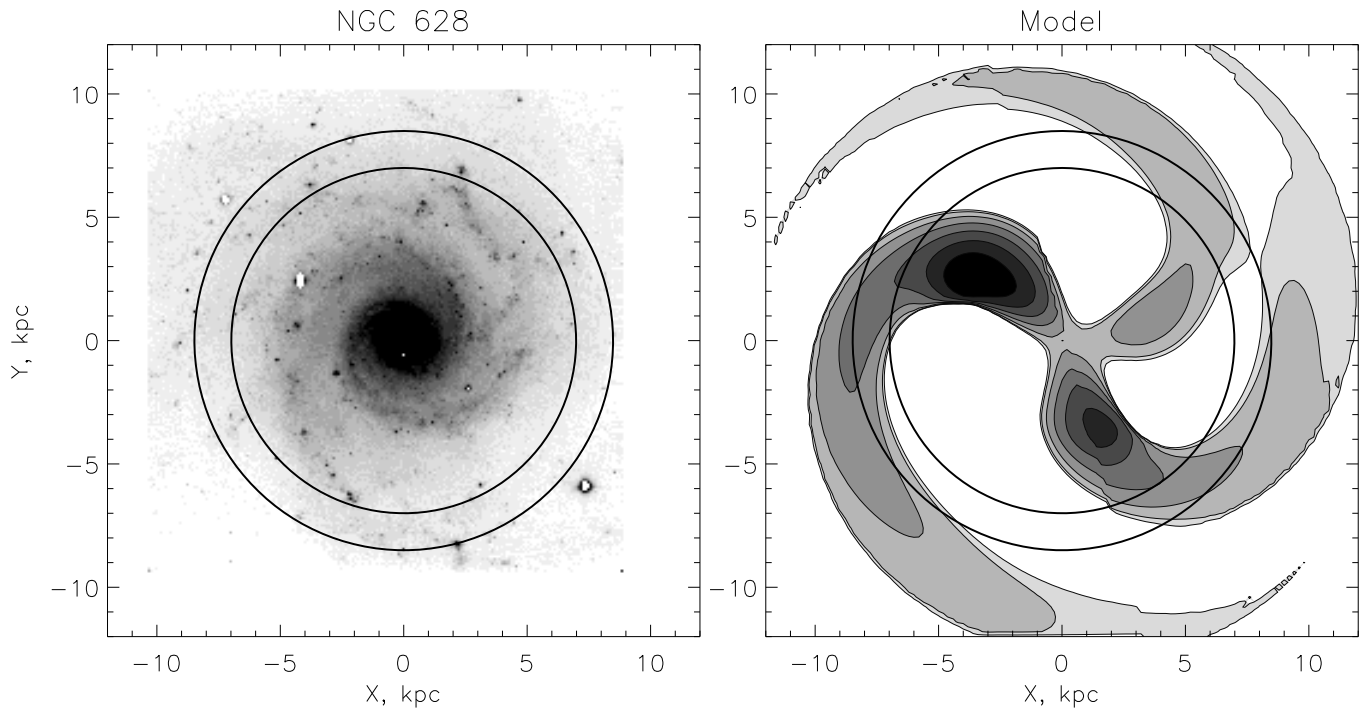}
  }}
  \caption{Left: a photograph in I-band of the galaxy NGC 628
(Larsen \& Richtler 1999). Right: the theoretical spiral pattern
taken at time 1.6$\times 10^9$ years from the beginning of
2-D simulations. The solid lines
show the positions of the corotation resonances for two-armed ($r_c=8.5$ kpc),
and three-armed spirals ($r_c=7.0$ kpc).
   \label{fig09}
   }
\end{figure*}

The rotation curve of NGC 628 was measured by Kamphuis \& Briggs (1992).
They find a maximum rotation velocity of the disk to be about 200 km s$^{-1}$ taking the
inclination of NGC 628 of 6.5\degr.
Figure 10 shows our best fit to the rotation curve of NGC 628 superimposed onto the
observational points from Kamphuis \& Briggs (1992).

With the inclination of NGC 628 of 6.5\degr, the observed data
provide direct measurement of the vertical component of the disk
velocity dispersion.
Van der Kruit \& Freeman (1984)
find the vertical velocity dispersion of the stellar disk to be
60 $\pm$ 20 \kms at a distance 60$^{\prime\prime}$ from the galactic center.
Following van der Kruit and Freeman (1984),
we assume that the velocity dispersion in the disk of NGC 628 exponentially
decreases with radius, and has a scale length twice larger than the
photometric scale length.
Van der Kruit \& Freeman (1984) find that the photometric scale length
of the old stellar disk in NGC 628  is about  60$^{\prime\prime}$.
Based on this value, the
kinematic scale length of the disk
of NGC 628 is equal to 120$^{\prime\prime}$.
This yields a value of the velocity dispersion at the disk center
equal to 99 $\pm$ 33 \kms.

\begin{figure}
   \resizebox{\hsize}{!}{
     \includegraphics[angle=-90]{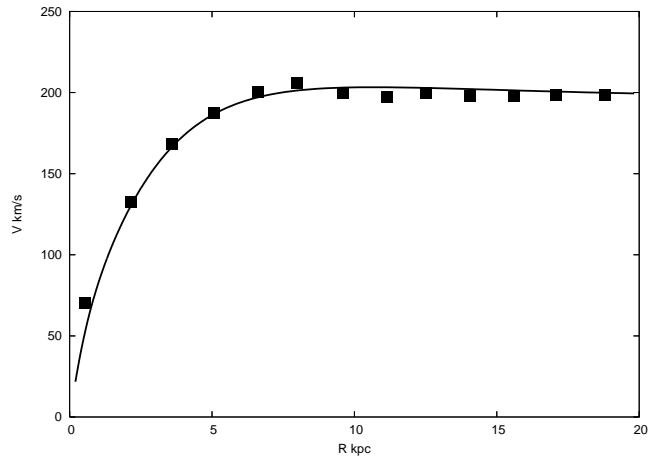}
   }
   \caption{Rotation curve of NGC 628. Observational points (filled squares)
are taken from Kamphuis and Briggs (1992). Solid line shows
the best fit to the observational points.
The parameters of the rotation curve modeled with the equation (3)
are: $\alpha =$ 0.5, $\beta =$ 0.5, $V_1 =$ 2.18, $V_2 =$ 0.47, $R_1 =$ 2.74,
$R_2 =$ 0.35. Units: $V =$ 148.5 km/s, $L =$ 2 kpc.
    }
   \label{fig10}
\end{figure}

Table 3 lists the parameters of the models studied in our linear
analysis. The vertical scale height of the disk varies in the models between
0.35 - 0.8 kpc, and the ratio of the vertical to the radial velocity dispersion
is allowed to vary within 0.5 - 0.8.
With these parameters, the disk mass changes from 4 $\times$ 10$^{10}$ M$_{\odot}$
to 9 $\times$ 10$^{10}$ M$_{\odot}$.

Linear analysis shows that the dynamics of perturbations in the disk
is not determined by one dominating mode.
We find that in most of the models the two global modes grow
simultaneously with comparable growth rates.
In the models with the lowest values of the velocity dispersion
allowed by observations, the three, and four-armed spirals are the most unstable
(models $a$ - $d$ in Table 3).
The models $e$ - $g$ that have the
most probable value of the central velocity dispersion of 99 km/sec,
are unstable towards $m=2$ and $m=3$ spirals. The models that  have the highest
possible velocity dispersion, support a two armed spiral prevailing
considerably over other competitors.

The models with the most probable velocity dispersion profile,
are unstable towards the two-armed, and the three-armed spirals.
These modes determine the dynamics of perturbations in the disk, and
the asymmetric shape of the spiral pattern in the disk of NGC 628
can be qualitatively explained by the interplay between these two dominant modes.

Nonlinear simulations are performed
for the model $e$ listed in Table 3.
This model has the velocity dispersion profile based on the measurements
by van der Kruit \& Freeman (1984) and the exponential surface density
distribution with the radial scale length of 2.9 kpc.
The vertical velocity dispersion profile is based on measurements
by by van der Kruit \& Freeman (1984), and has
kinematical scale length of 5.8 kpc.
The total mass of the disk
is about 6$\times$ 10$^{10}$ M$_{\odot}$, and the ratio of the velocity dispersions
$c_z/c_r$ is 0.6.

\begin{figure}
   \resizebox{\hsize}{!}{
     \includegraphics[angle=-90]{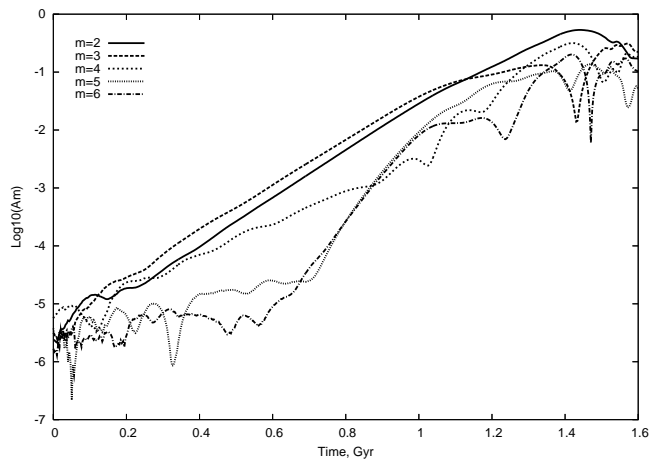}
   }
   \caption{Global Fourier amplitudes calculated for perturbations developing in the model
$e$ of galaxy NGC 628. The two-armed and the three-armed spirals grow with the
approximately equal growth rates.
    }
   \label{fig11}
\end{figure}

Figure 11 shows the time dependence of the global amplitudes
for spiral modes $ m= 2 - 6$ calculated with help of equationb (5).
The two-armed, and three-armed
modes grow with the approximately equal growth rates.
At the beginning, the three-armed spiral
slightly dominates the dynamics of the perturbations. Later on, the two-armed spiral
reaches a comparable amplitude, which  results in an asymmetric
two-armed spiral pattern.
The right panel of Figure 9 shows a theoretical contour plot of the density distribution
in the disk of NGC 628 taken at time 1.6 $\times$ 10$^9$ years.
The Figure shows also the positions of corotation resonances for two
governing modes, $m=2$ with $r_c=8.5$ kpc, and $m=3$ global mode with $r_c=7.0$ kpc.
As it can be seen from the Figure, the basic features of spiral structure of NGC 628,
namely the two-armed spiral pattern with a third arm of comparable amplitude,
are reproduced in our simulations.

%%%%%%%%%%%%%%%%%%%%NGC 2985%%%%%%%%%%%%%%%%%%%%%%%%

\subsection{ NGC 2985}

The Sab galaxy NGC 2985 is another example of
an early type galaxy in our sample.
This galaxy has an inclination of about 36$\degr$ (Grosbol 1985),
and distance of 18 Mpc with Hubble constant of 75 km/s/Mpc.
The spiral pattern in NGC 2985 is
similar to that in galaxy NGC 488.
NGC 2985 has a regular two-armed
spiral pattern outlined by tightly wound narrow arms, as can be seen on
the left panel of Figure 12. The Figure reproduces high-resolution
R-band image of NGC 2985 taken from the NED archive (Knapen et al. 2004).
Similarly to the image of NGC 488, the bulge component has been subtracted
assuming a Sersic (1968) profile for the bulge ($I(r)=I_0exp(-(r/r_0)^{1/n})$,
$n=3$), and an exponential surface brightness distribution for the disk
of NGC 2985.

\begin{figure*}[tp]
  \centerline{\hbox{
  \includegraphics[width=\textwidth]{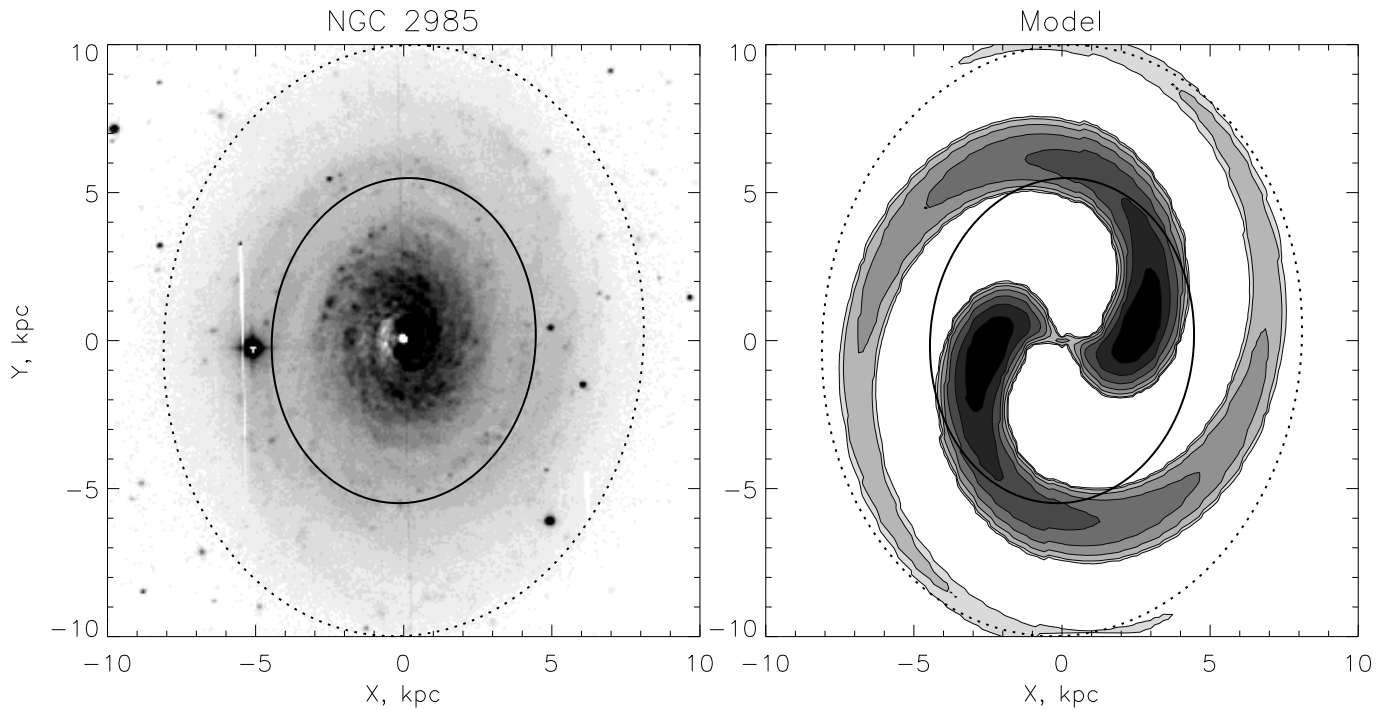}
  }}
  \caption{Left: a photograph in I-band of the galaxy NGC 628
(Larsen \& Richtler 1999). Right: the theoretical spiral pattern
taken at time 1.6$\times 10^9$ years from the beginning of
2-D simulations. The solid lines
show the positions of the corotation resonances for two-armed ($r_c=8.5$ kpc),
and three-armed spirals ($r_c=7.0$ kpc).
   \label{fig12}
   }
\end{figure*}

We build equilibrium models of NGC 2985 based
on measurements by Gerssen et al. (2000).
Figure 13 shows the observed rotation curve of the stellar disk
of NGC 2985 obtained with long-slit
absorption spectra along the major axis of the galaxy. The dashed line
in Figure 13 is our best-fit to the observational data used in the
numerical simulations.

\begin{figure}
   \resizebox{\hsize}{!}{
     \includegraphics[angle=-90]{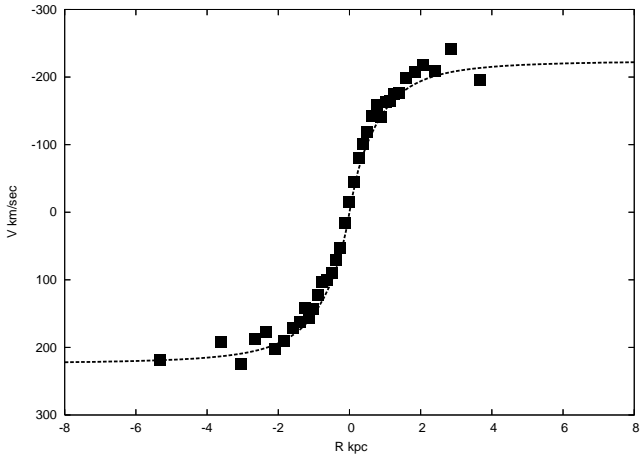}
   }
   \caption{Global Fourier amplitudes calculated for perturbations developing in the model
$e$ of galaxy NGC 628. The two-armed and the three-armed spirals grow with the
approximately equal growth rates.
    }
   \label{fig13}
\end{figure}

NGC 2985 is one of the few galaxies that have a measured ellipsoid
of the stellar velocity dispersions.
Gerssen et al. (2000)
find the values of the radial and vertical velocity dispersions
at the disk center of
149$\pm$12 and 127 $\pm$ 10 \kms correspondingly, and the ratio
of the vertical to the radial velocity dispersion in the disk of NGC 2985
of $0.85\pm0.1$.
The radial scale length
of the exponentially decreasing velocity dispersions
from the Gerssen et al. (2000) measurements is about  $73\pm 9^ {\prime\prime}$.
This value is approximately two times
larger than the value of the photometric scale length  in I band
of $30^{\prime\prime} \pm 4 ^{\prime\prime}$.
We use the value of $30^{\prime\prime}$ for the radial scale length
of the exponentially decreasing surface density distribution.
Some models are built using
the radial scale length of the surface density distribution of $36^{\prime\prime}$
which is based on the kinematical scale length obtained by Gerssen et al. (2000).
The results from these two sets of models are comparable.

As in the case of NGC 488, we have studied a family of models
varying the vertical scale height of the disk within 0.5 - 0.8 kpc,
and the vertical velocity dispersion within the observational errors $\pm$ 10 \kms.
Table 4 summarizes the equilibrium and the stability properties of the models.
In accordance with measurements, the stellar radial velocity dispersion
at the disk center varies in the models within 137 - 161 km/s.
The surface density at the disk center changes
within 1.25$\times$ 10$^9$ - 1.75$\times$ 10$^9$
M$_{\odot}$ kpc$^{-1}$.

Right panel of Figure 12 shows the
results of our nonlinear simulations for one of the models of the galaxy NGC 2985.
In this particular model, the radial scale length of the surface density distribution
was chosen to be 3.1 kpc ( $36^{\prime\prime}$).
The vertical and the radial velocity dispersions
are taken 137 km/s and 117 km/s respectively,
and the vertical scale height of the disk density distribution is
equal to 0.8 kpc. With these parameters, the total mass of the
disk is 7.14 $\times$ 10$^{14}$ M$_{\odot}$.
The simulations start
from noise perturbations of amplitute of order of 10$^{-15}$.
We find that the initial amplitude of perturbations
does not affect the properties of an emerging spiral pattern,
and determines only a duration of a linear exponential growth phase.

Soon after the beginning of simulations,
$m=2$ global spiral mode emerges from the arbitrary noise,
and prevails over other unstable modes until the nonlinear saturation
phase is reached. The right panel of Figure 12 shows a two-dimensional contour plot
of the density distribution taken at time 2.74 $\times$ 10$^9$ years from the
beginning of simulations.
The theoretical pattern shows good qualitative agreement with the
observed spiral arms in the galaxy NGC 2985.
In the inner regions, the theoretical arms are more open
compared to the observed ones.
A disagreement between the observed and the theoretical
spiral patterns in the inner parts of the galactic disk
can be probably be attributed to the uncertainty of our knowledge of
the disk properties within the bulge region.
Within all observational errors, however,
the two-armed spiral pattern prevails over other unstable modes,
and determines the properties of perturbations in the disk
of NGC 2985.

\subsection {NGC 3938}

NGC 3938 is an Sc  galaxy that has
an inclination of about $14\degr$, and distance of 11.3 Mpc  (Jim\'enez-Vicente et al. 1999).
The left panel of Figure 14 shows the I-band image of NGC 3938 obtained with the
Mauna Kea 2.2m telescope (Tully et al. 1996), taken from the NED archive.
As can be seen from the Figure, NGC 3938 is a typical multi-armed spiral
with the spiral arms branching from its center.

The rotation curve and
the velocity dispersion of NGC 3938 have been measured, allowing us to build
an equilibrium model for this galaxy.
In Figure 15 is plotted the rotation curve of NGC 3938 taken from Jim\'enez-Vicente et al. (1999).
A solid line shows
our best fit to the observational data we use in our numerical simulations.

\begin{figure*}[tp]
  \centerline{\hbox{
  \includegraphics[width=\textwidth]{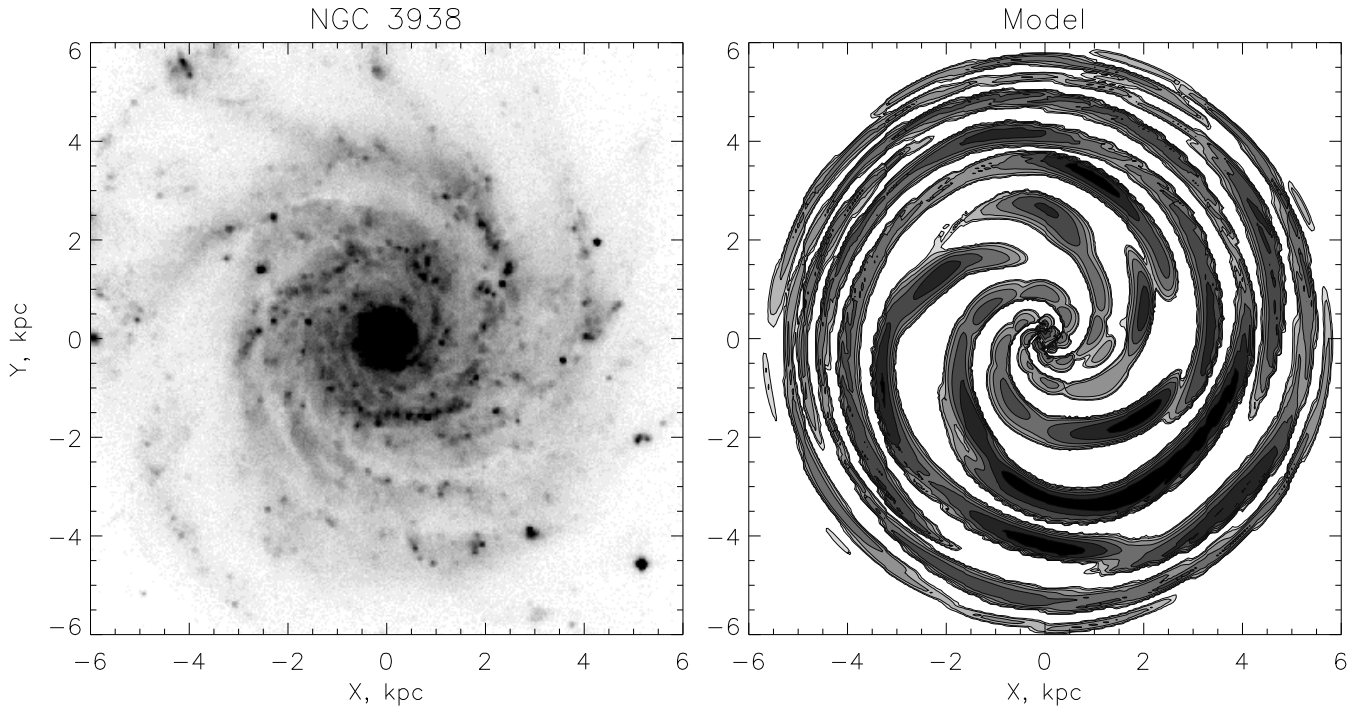}
  }}
  \caption{On the left panel is shown I-band image of NGC 3938 taken from the NED archive
(Tully et al. 1996). On the right panel is shown the snapshot from 2D simulations
of model $e$, taken at time 0.7 $\times$ 10$^9$ years.
   \label{fig14}
   }
\end{figure*}

The rotation curve and
the velocity dispersion of NGC 3938 have been measured, allowing us to build
an equilibrium model for this galaxy.
In Figure 15 is plotted the rotation curve of NGC 3938 taken from Jim\'enez-Vicente et al. (1999).
A solid line shows
our best fit to the observational data we use in our numerical simulations.

The stellar velocity dispersion profile of NGC 3938 was
measured by Bottema (1988). Due to the small inclination of NGC 3938,
the line of sight velocity dispersion is determined mainly by the component of
the velocity dispersion perpendicular to the disk.
Bottema (1988) finds that the z-component of the velocity
dispersion can be described by an exponential function
with a velocity dispersion at the disk center  of 35$\pm$5 \kms,
and a scale length of 3.3 kpc. The velocity dispersion measurements
of Bottema (1998) together with the velocity dispersion profiles
used in our models are also plotted in Figure 15.

\begin{figure}
   \resizebox{\hsize}{!}{
     \includegraphics[angle=-90]{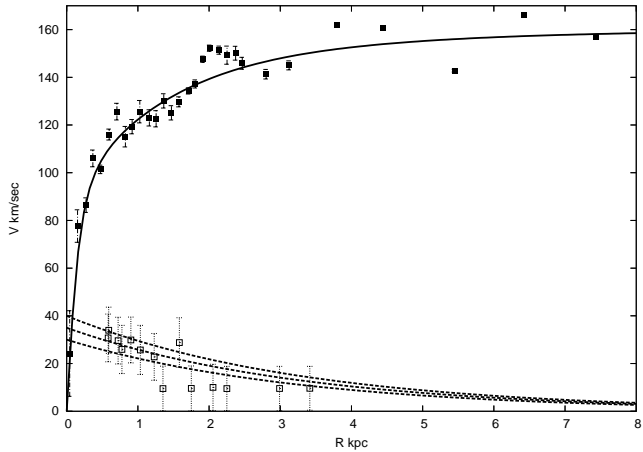}
   }
   \caption{Filled squares with error bars show the rotation curve of NGC 3938.
The data points are taken from Jim\'enez-Vicente et al. (1999). Solid line is
the best fit to the observational points.
The parameters of the rotation curve modeled with the equation (3) are:
$\alpha =$ 0.5, $\beta =$ 0.5, $V_1 =$ 0.41, $V_2 =$ 0.68, $R_1 =$ 1.17,
$R_2 =$ 0.1. Units: $V =$ 148.5 km/s, $L =$ 2 kpc.
Dashed lines - the velocity dispersion profiles studied in our models of NGC 3938.
The radial scale length of 3.5 kpc, and the values of the velocity dispersion at the disk center
of 30, 35 and 40 km/sec are assumed.
The open squares with error bars are the observational points for the velocity
dispersion taken from Bottema (1988).
    }
   \label{fig15}
\end{figure}

The thickness of the disk of nearly face-on NGC 3938 can be estimated
using velocity dispersion measurements. In a rotating self-gravitating
disk in equilibrium, the disk vertical scale height, $z_0$, is related to
the maximum rotation velocity $v_{max}$ and to the radial scale length of
the exponential density distribution, $h$, as (Freeman 1970):
\be
    v_{max} = 0.88 \sigma_z(R=0) (h/z_0)^{1/2}
   \label{eq8}
\ee

The disks of real galaxies are supported by the combined disk self-gravity
and halo potential. Bottema (1997) estimates that the disk of NGC 3938 can account for
about 60 percent of the maximum rotational velocity of the disk.
With this assumption, and taking the velocity dispersion at the disk
center equal to 35 km/s and the radial scale length of the disk density
distribution of 1.65 kpc, we find that the vertical scale height of the
disk density distribution in NGC 3938
is about 0.2 kpc. We use models with a vertical scale height between
0.125 and 0.25 kpc. We find that with the scale heights larger than
0.25 kpc the disk of NGC 3938 is stable against spiral perturbations.

Table 5 summarizes the studied models and their stability properties.
The linear stability analysis shows that the models
are the most unstable towards the $m=3$ and $m=4$ global
modes, which grow with the comparable rates. The
superposition of the three-armed and the four-armed
spirals of approximately equal amplitudes determines the multi-armed appearance of
NGC 3938.

The nonlinear simulations confirm this conclusion.
Figure 16 shows the time dependence of the global amplitudes
for $m=1-4$ spirals calculated for model $e$ in Table 5.
The simulation starts with the equilibrium disk seeded
with noise perturbations with an amplitude of 10$^{-6}$
of the equilibrium disk surface density.
In agreement with the linear analysis, the disk is most
unstable towards $m=3$ and $m=4$ spirals with the four-armed spiral slightly dominating
over other competitors. The right panel of Figure 14
shows a snapshot of our nonlinear simulation taken at
0.7 $\times$ 10$^9$ years.
As one can see, the disk indeed generates a multi-armed spiral pattern qualitatively
resembling the observed spiral pattern.

\begin{figure}
   \resizebox{\hsize}{!}{
     \includegraphics[angle=-90]{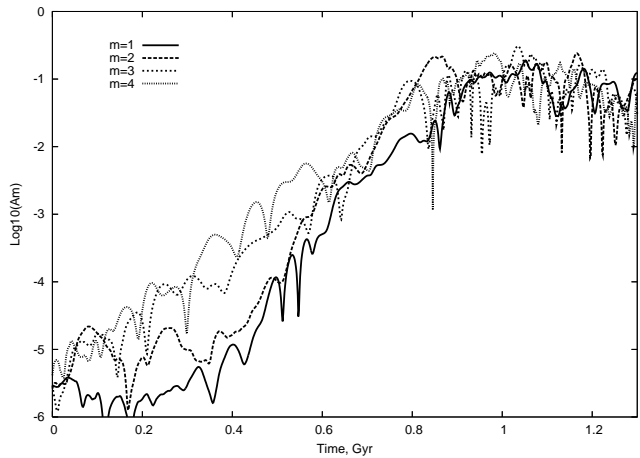}
   }
   \caption{Global Fourier amplitudes developing in model $e$ of the galaxy NGC 3938.
    }
   \label{fig16}
\end{figure}

As in the most of our hydrodynamical simulations, soon after the nonlinear saturation stage
(about 0.9 Gyr for the model $e$), the spiral pattern re-organises itself, and the spiral
arms are usually destroyed in the central regions of the disk.
Typically, the mass concentration grows near the center, and the spiral pattern
fragments into a few orbiting clumps.
Figure 17 shows the evolution of the density perturbations in our model $e$
during approximately 2 Gyr.
During the exponential growth phase, that lasts
about 0.9 Gyr, and at the beginning of the nonlinear saturation phase,
the spiral pattern keeps its multi-armed nature. However, in the outer regions of the disk,
the spiral armlets can be seen at the later stages of evolution as well
(see, e.g., the last two frames of Figure 17). These armlets are
associated with the density clumps, and can be a result of swing
amplified reaction of the disk to the perturbations imposed by the
clumps' gravitational force.

\begin{figure}
   \resizebox{\hsize}{!}{
     \includegraphics[angle=0]{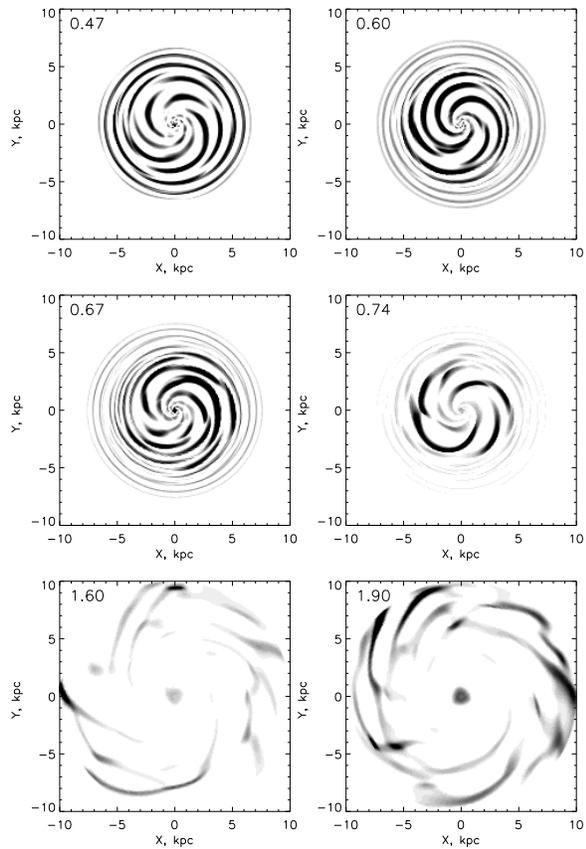}
   }
   \caption{A long-tem evolution of the density perturbations developing in the model $e$ of NGC 3938.
At the nonlinear saturation stage, the spiral pattern fragments into a few clumps
and spiral armlets seen in the outer regions of the disk. Time is given in units of 10$^9$ years.
    }
   \label{fig17}
\end{figure}

\subsection {NGC 6503}

NGC 6503 is a highly inclined (inclination of 74$\degr$) late-type spiral Sc galaxy,
which is the nearest object in our sample at a distance of 6 Mpc (Bottema 1989).
The rotation of NGC 6503 has been studied by a number of authors.
We use the HI rotation curve taken from Begeman et al. (1991).
The observed rotation curve was fit to equation (3), and both the observations
and fit (Figure 18) show good agreement with the measurements of Karachentsev \& Petit, (1990),
and de Vaucouleurs \& Caulet(1982).
The rotation velocity
reaches a maximum value of about 120 km s$^{-1}$, and remains nearly constant
in the outer regions of the disk.

The stellar velocity dispersion in NGC 6503 was studied by Bottema (1989).
The orientation of NGC 6503 enables one to measure the transverse components of the velocity
dispersion in the disk.
Bottema's measurements yield a value for the radial component
of the velocity dispersion at the disk center of 55 $\pm$ 10 \kms,
and a kinematical scale length of 80$^{\prime\prime}$ - twice as large as the photometric
scale length of the disk.
The observed line-of-sight velocity dispersion of stellar disk in NGC 6503,
together with the exponential fits to the physical velocity dispersion in R-direction
are shown in Figure 18.

\begin{figure}
   \resizebox{\hsize}{!}{
     \includegraphics[angle=0]{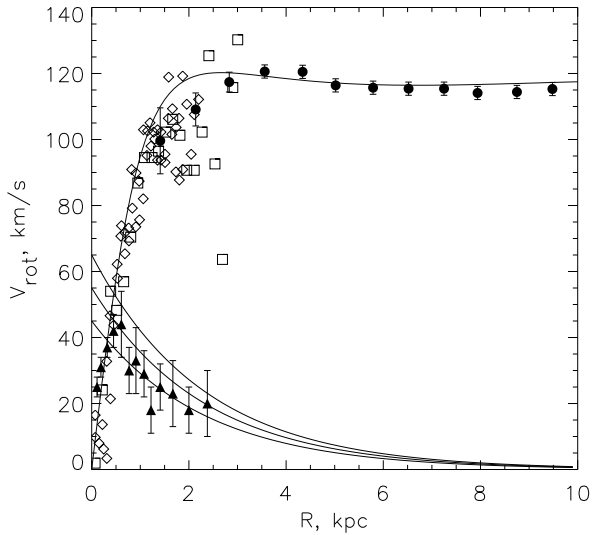}
   }
   \caption{Rotation curve of the galaxy NGC 6503. Filled circles - HI observational data of
Begeman et al. (1991). Open squares - measurements of Vaucouleurs and  Caulet (1982),
open diamonds - measurements of Karachentsev \& Petit (1990). The solid line -
the best fit to observational data. Parameters of the rotation curve:
$\alpha =$ 0.75, $\beta =$ 0.75, $V_1 =$ 480 km/s/kpc$^{1/2}$, $V_2 =$ 202 km/s/kpc$^{1/2}$, $R_1 =$ 4.12 kpc,
$R_2 =$ 1.19 kpc.
Filled triangles with error bars show the observed velocity dispersion of stars
in the disk of NGC 6503 taken from Bottema (1989).Superimposed are exponential velocity
dispersion distributions with the radial scale length of 2.3 kpc,
and the values of the velocity dispersion at the disk center of 45, 55 and 65 km/sec.
    }
   \label{fig18}
\end{figure}

A peculiar feature of NGC 6503 is an abrupt drop of the
disk velocity dispersion near its center, which
Bottema \& Gerritsen (1997) speculate is
related to a component that is  kinematically distinct from the disk.
Another peculiarity is the existence of
two exponential scale lengths in the surface brightness distribution of its disk.
In the inner regions ($R<160^{\prime\prime}$),
the scale length is about 40$^{\prime\prime}$, while in the outer regions
of the disk the scale length is about  80$^{\prime\prime}$ (Bottema 1989).

Using the rotation curve and the velocity dispersion profile along the radius
of the disk, we have built  a set of equilibrium models for NGC 6503.
As in the previous cases, we consider locally stable disks
with Toomre Q-parameter Q$_{min} >$ 1.0.
The models, listed in Table 6, are the disks with an
exponential surface density distribution that have a scale length of 1.16 kpc (40$^{\prime\prime}$)
and a kinematical  radial scale length equal to 2.32 kpc (80$^{\prime\prime}$).
The radial component of the velocity dispersion at the disk center was taken to be  55 $\pm$ 10 \kms,
which yields a total mass for the disk of 4 - 7 10$^9$ M$_{\odot}$.
To explore the peculiar properties of NGC 6503, we also studied a model that has
two exponential scale lengths in the disk density distribution,
and a model with a drop of the velocity dispersion near the disk center.
Remarkably, all models demonstrate a similar  behavior.
becoming unstable towards the
$m=3$, and $m=4$ modes which have comparable growth rates, resulting
in a multi-armed appearance of spiral pattern.

Nonlinear modeling confirms this conclusion.
In agreement with the linear analysis, the time dependence of the global Fourier
amplitudes presented in Figure 19 shows that the
disk is most unstable  towards $m=4$ global mode.
Later on, $m=3$ harmonics as well as the global models with a higher
number of arms reach a comparable  amplitude, which
results in a multi-armed appearance of the spiral pattern
observed in the disk of NGC 6503.
The right panel of 20 shows the results of the nonlinear simulations for model "d".
In this model, a velocity dispersion at the disk center of
55 km/sec was adopted, the ratio of the vertical to the radial velocity dispersions
$c_z/c_r$ was taken as 0.7, and the vertical scale height $z_0$ of the disk density
distribution was taken as 0.17 kpc. The total mass of the disk is then about
5.2 $\times$ 10$^9$ M$_{\odot}$.
The surface density contour plot taken at time 10$^9$ years
reproduces qualitatively well the observed multi-armed morphology of the
galaxy.

\begin{figure}
   \resizebox{\hsize}{!}{
     \includegraphics[angle=-90]{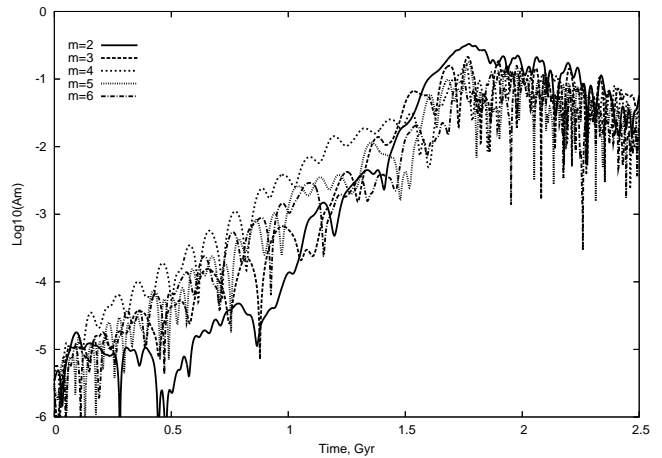}
   }
   \caption{Global Fourier amplitudes developing in model $d$ of the galaxy NGC 6503.
The $m=3$ -- $m=6$ global modes grow with the comparable growth rates.
    }
   \label{fig19}
\end{figure}

\begin{figure*}[tp]
  \centerline{\hbox{
  \includegraphics[width=\textwidth]{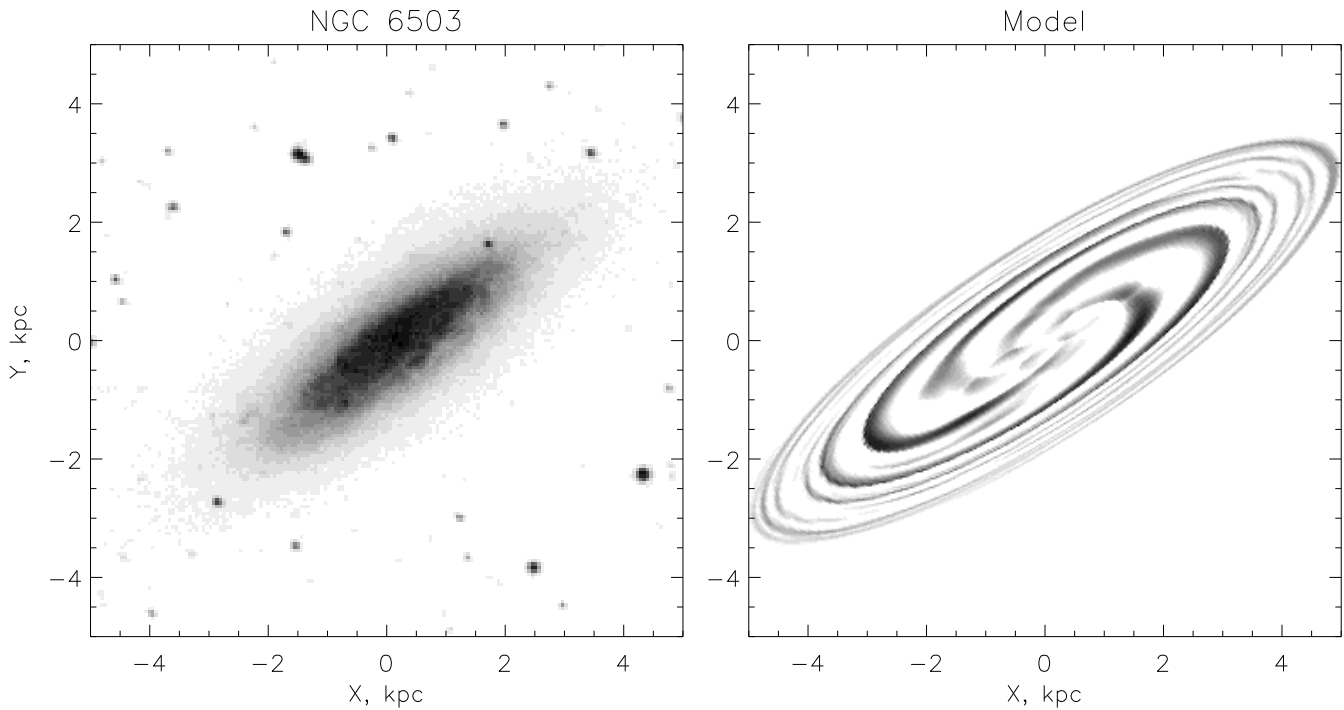}
  }}
  \caption{Left: a POSS2 R-band image of NGC 6503. Right:
the contour plot for the perturbed density distribution taken
from numerical 2D simulations at time
$10^{9}$ years.
   \label{fig20}
   }
\end{figure*}

%scussion%%%%%%%%%%%%%%%%%%%%%%%%%%%%%%%%%%%%%%%%%%%%%%%%%%%%

\section{Summary and Discussion}

We have studied the applicability of the global modes
as an explanation for the spiral structure of disk galaxies.
We selected six nearby spiral galaxies with measured axisymmetric background properties for their disks.
Using the observed radial distributions for the stellar velocity dispersions
and the disk rotation velocities, we constructed equilibrium properties
for the galactic disks in each galaxy and implemented two independent
theoretical analyses of the spiral structure in the disks.
Using linear global modal analysis, we determine the set
of unstable global modes existing in the disk of each particular galaxy.
We then use 2-D nonlinear hydrodynamical simulations to model the dynamics
of the galactic disks seeded initially with noise perturbations which
have relative amplitudes on the order of 10$^{-6}$ -- 10$^{-12}$.
The net result of our work is that both kinds of theoretical analyses
predict spiral patterns in the disks of the particular galaxies which
agree qualitatively  with the observed morphological properties
of spirals.

We find that the overall morphological properties of the
spiral patterns are largely insensitive to the details
of the models, such as the thickness of the galactic disks, or the ratio
of the velocity dispersions within observational error limits.
Massive disks with a high stellar velocity dispersion
in their central regions  have a tendency to support
two-armed spirals.  Examples are the galaxies
NGC 1566, NGC 488 and NGC 2985.
The disk of NGC 628 has somewhat 'intermediate' properties with the two-armed and three-armed
spiral modes having comparable growth rates.
The velocity dispersion in the central region of NGC 628 is less than 100 \kms,
and its evolution is determined by a tight competition of the two-armed, and three-armed spirals.
Two Sc galaxies in our sample, NGC 3938 and NGC 6503, have low-mass disks and low  velocity dispersions,
which result in a multi-armed spiral structure. This agrees with the conclusion
of Evans $\&$ Read (1998) who
found that the three-armed or four-armed spirals  occur only in disks with low velocity dispersion.

The theoretical spiral pattern in Sab galaxy NGC 2985, and especially in Sb galaxy NGC 488 is more
open when compared to the observed spirals. The reason for this might be in the uncertainties
related to the bulge-disk decomposition, and to a contribution of the bulge component
to the dynamics of spiral patterns. However, the overall morphology
of the spiral patterns in NGC 488 and NGC 2985 is correctly predicted by the global modal approach.

An over-reflection of the spiral density waves,
i.e. transmission and larger amplitude reflection at the corotation resonance
of the spiral wave trains, is seen by Lin and his collaborators as a physical explanation
of the excitation mechanism for the global modes in gravitating disks
(Lau et al. 1976). From our numerical simulations, it is hard
to depict the radially propagating wave trains that "over-reflect"
at the corotation resonance. Nevetheless, we calculated an amplification factor
of spirals based on this picture. Drury (1980) developed an analytical method
to calculate the reflection and transmission coefficients for the global modes in an unstable
shearing gas sheet. The amplification factor was calculated
for $m=2$, $m=3$ and $m=4$
spiral modes for all six galaxies in our sample.
Usually, a dominant spiral mode has the highest amplification factor.
The amplification factor for the $m=2$ spiral mode is largest in
the disks of the two-armed spiral galaxies NGC 488, NGC 1566 and NGC 2985.
In the multi-armed spiral galaxies NGC 3938 and NGC 6503, the amplification factors for
$m=4$, and $m=3$  modes are dominant.
This result can be considered as a hint that the amplification at the corotation
resonance might indeed be at work in growing spiral modes.

According to Toomre (1981), the essence of the growing modes
can be understood in terms of the swing amplification.
For a $V=$ constant disks, the swing amplification is strongest for
a dimensionless `unwrapped' wavelength $X = \lambda_t / \lambda_c$
of about 1.5 (Toomre 1981).  In this model,
$$
  X = {2 \over m} (1 +f)
$$
where m is a number of arms, and $f$ is a ratio of halo radial force
to the unperturbed disk force (Binney \& Tremaine 1987).
In the low-mass disk,  the ratio $f$ is large, and the multi-armed spirals are swing-amplified.
This is in a qualitative agreement with our results
showing that the low-mass disks with low velocity dispersion
support multi-armed spiral pattrens.
A further study, however, is needed to make a conclusion as to
whether the growth of the spiral global modes is
explained by a swing amplification mechanism.

What can be concluded from our simulations as to whether the spiral patterns in galaxies
are a long-lived or a short-lived phenomenon,  regenerated many times
during the galactic evolution? The theoretical spiral patterns make
typically about ten revolutions in our simulations. Usually, after the nonlinear
saturated stage of instability starts, the modes grow to sufficiently large
amplitudes, and shocks occur. The shocks dissipate energy and eventually
destroy the spiral pattern. At this highly nonlinear stage, the hydrodynamical
approximation is inadequate  for the description of the collisionless stellar disks.
The is one example, the galaxy NGC 1566,
in which the nonlinear coupling of the modes stabilizes an exponential growth
before the pattern disruption by  shocks occurs, and
the two-armed spiral pattern exists throughout the nonlinear
saturated stage.

A duration of the exponential growth of a spiral pattern until it is
disrupted by the nonlinear effects is about one - two Gyr.
This gives an upper limit on the number of such regenerations
to be about of a few during the galactic evolution.
The example of NGC 1566 shows however, that the spiral pattern
can exist during a nonlinear saturation phase as well,
indicating that the spiral structure in galactic disks can indeed be
a long-lived phenomenon.

Further study of the dynamics of collisionless disks is of principal importance
in clarifying this question.

%%%%%%%%%%Acknowledgments%%%%%%%%%%%%%%%%%%%%%%%%%%%%%%

\begin{acknowledgements}

The authors thank S.Shore, A. Toomre and J. Kenney for valuable comments,
and J. Gerssen and K. Kuijken for providing data on NGC 488 and NGC 2985.
The authors thank also T.M. Girard, W.F. van Altena, D. Maitra and R. Quadri for reading the manuscript,
which resulted in in mproved presentation.
A. Moiseev thanks the Russian Science Support Foundation.
This research has made use of the NASA/IPAC
Extragalactic Database (NED) which is operated by the Jet
Propulsion Laboratory, California Institute of Technology, under contract
with the National Aeronautics and Space Administration.
Some of the data presented in this paper were obtained from the
Multimission Archive at the Space Telescope Science Institute (MAST). STScI is
operated by the Association of Universities for Research in Astronomy, Inc.,
under NASA contract NAS5-26555. Support for MAST for non-HST data is provided
by the NASA Office of Space Science via grant NAG5-7584 and by other grants
and contracts.

\end{acknowledgements}

%%% Table 1 %%%
\begin{table*}
\caption{The sequence of equilibrium models, and the parameters of the most unstable $m=2$
global mode for galaxy NGC 1566. Columns give the vertical velocity dispersion at the disk center, $c_z$,
the ratio of the velocity dispersions $c_z/c_r$, the vertical scale height of the disk $z_0$, the surface
density of the disk at the disk center $\sigma_0$, the total mass of the disk $M_{disk}$, and
the minimum value of Q-parameter.ЕThe last column gives the values of the pattern speed and the growth
rate  of the unstable mode.
}
\label{table_1}
\centering
\begin{tabular}{cccccccc}
\hline\hline
Model & $c_z$ & $c_z/c_r$ & $z_0$ & $\sigma_0$ & $M_{disk}$ & $Q_{min}$ & ($\Omega_p$, Im$\omega$)   \\
      &  (km s$^{-1}$)&          & (kpc) & ($10^9 M_{\odot}/ kpc^2$)& ($10^{10} M_{\odot}$) &  & (\kmskpc) \\
\hline
{\em a} & 120 & 0.5  &  0.30 & 3.43 & 3.64 & 1.49 & (42.6, 7.9) \\
{\em b} & 120 & 0.6  &  0.35 & 2.94 & 3.12 & 1.44 & (49.5, 9.6) \\
{\em c} & 120 & 0.7  &  0.45  & 2.29 & 2.43 & 1.58 &(46.2, 4.1) \\
{\em d} & 120 & 0.8  &  0.50   & 2.06 & 2.18 & 1.55 & (50.9, 8.8) \\
{\em e} & 148 & 0.5  & 0.40 & 3.92 & 4.15 & 1.60 & (32.8, 3.7) \\
{\em f} & 148 & 0.6  & 0.45 & 3.48 & 3.70 & 1.51 & (41.3, 7.3) \\
{\em g} & 148 & 0.7  & 0.55 & 2.85 & 3.02 & 1.57 & (42.8, 5.4) \\
{\em h} & 148 & 0.8  & 0.60 & 2.61 & 2.77 & 1.52 & (48.6, 14.2) \\
{\em i} & 176 & 0.5  & 0.45 & 4.93 & 5.22 & 1.50 & (31.1, 4.0)  \\
{\em j} & 176 & 0.6  & 0.50 & 4.43 & 4.70 & 1.42 & (40.2, 10.0) \\
{\em k} & 176 & 0.7  & 0.60 & 3.69 & 3.91 & 1.44 & (42.9, 9.2) \\
{\em l} & 176 & 0.8  & 0.75 & 3.02 & 3.20 & 1.47 & (48.9, 12.5) \\
\hline
\end{tabular}
\end{table*}

%%% Table 2 %%%

\begin{table*}
\caption{Equilibrium models of the galaxy NGC 488.
The columns give the radial velocity dispersion at the disk center,
$c_r$, the surface density at the center of the disk, $\sigma_0$,
the total mass of the disk, $M_{disk}$, the minimum value of
Q-parameter. The pattern speed $\Omega_p$ and the growth rate Im$\omega$
of the most unstable $m=2$ mode are given in the last column
of the Table.}
\label{table_2}
\centering
\begin{tabular}{cccccc}
\hline
\hline
Model & $c_r$ & $\sigma_0$ & $M_{disk}$ & $Q_{min}$ & $(\Omega_p$, Im$\omega)$   \\
      & (km s$^{-1}$)  & ($10^9 M_{\odot} / kpc^2$) & ($10^{10} M_{\odot}$) &  & (\kmskpc)  \\
\hline
{\em a} & 285 & 2.6 & 10.9 & 1.7 & stable \\
{\em b} & 285 & 3.0 & 13.5 & 1.45 & (52.5, 14.7) \\
{\em c} & 285 & 3.2 & 14.6 & 1.34 & (57.0, 19.4) \\
{\em d} & 285 & 3.5 & 15.9 & 1.23 & (62.2, 25.3) \\
{\em e} & 285 & 3.9 & 17.5 & 1.11 & (68.4, 32.8) \\
{\em f} & 285 & 4.1 & 18.2 & $<$ 1 &  \\
{\em g} & 253 & 2.3 & 10.2 & 1.7 & stable \\
{\em h} & 253 & 2.6 & 11.7 & 1.5 & (55.2, 11.9) \\
{\em i} & 253 & 3.0 & 13.5 & 1.3 & (63.4, 19.2) \\
{\em j} & 253 & 3.2 & 14.6 & 1.2 & (67.8, 24.4) \\
{\em k} & 253 & 3.5 & 15.9 & 1.1 & (74.9, 31.7) \\
{\em l} & 253 & 3.8 & 16.9 & $<$ 1 &  \\
{\em m} & 221 & 2.1 &  9.3 & 1.65 & stable  \\
{\em n} & 221 & 2.4 & 11.0 & 1.5 & (61.8, 12.1)  \\
{\em o} & 221 & 2.6 & 11.7 & 1.32 & (65.9, 15.4)  \\
{\em p} & 221 & 2.8 & 12.5 & 1.24 & (70.1, 19.1)  \\
{\em q} & 221 & 3.0 & 13.5 & 1.15 & (75.0, 23.8)  \\
{\em r} & 221 & 3.2 & 14.4 & $<$ 1 &   \\
\hline
\end{tabular}
\end{table*}

%%% Table 3 %%%

\begin{table*}
\caption{The sequence of equilibrium models and the parameters of
$m=2$, $m=3$, and $m=4$ global modes growing in the studied models of NGC 628.
The last column gives the number of arms, the pattern speeds and the growth rates
of the modes growing in a particular model.}
\label{table_3}
\centering
\begin{tabular}{cccccccc}
\hline
\hline
Model & $c_z$ & $c_z/c_r$ &$z_0$ & $\sigma_0$ & $M_{disk}$ &$Q_{min}$ & $(\Omega_p$, Im$\omega)$   \\
  & (km s$^{-1}$) &  & (kpc) & ($10^9 M_{\odot / kpc^2}/ kpc^2$) & ($10^{10} M_{\odot}$) &  & (\kmskpc)  \\
\hline
{\em a} & 70 & 0.5  &  0.35  & 1.0 &  5.1 & 1.31 &3 (30.9, 8.6) \\
{\em b} & 70 & 0.6  &  0.4   & 0.9 & 4.5 & 1.25 &3 (35.5, 9.4) {\mbox ;}~4 (36.5, 8.2) \\
{\em c} & 70 & 0.7  &  0.45  & 0.8 & 4.0 & 1.2  &3 (38.5, 9.2) {\mbox ;}~4 (40.1, 8.9) \\
{\em d} & 70 & 0.8  &  0.5   & 0.7 & 3.3 & 1.17 &3 (41.1, 8.5) {\mbox ;}~4 (43.1, 9.2) \\
{\em e} & 99 & 0.5 & 0.5    & 1.4 & 7.3 & 1.31  &2 (21.9, 12.0){\mbox ;}~3 (25.8, 9.7) \\
{\em f} & 99 & 0.6 & 0.6    & 1.2 & 6.0 & 1.31  &2 (23.4, 10.1){\mbox ;}~3 (28.5, 9.4) \\
{\em g} & 99 & 0.7 & 0.7    & 1.0 & 5.2 & 1.32  &2 (24.2,  7.7){\mbox ;}~3 (30.6, 8.8) \\
{\em h} & 99 & 0.8 & 0.75   & 0.95 & 4.8 & 1.23  &3 (34.5, 10.8){\mbox ;}~4 (35.4, 9.1) \\
{\em i} & 130 & 0.5 & 0.65   & 1.9 & 9.4 & 1.31 &2 (19.4, 12.6){\mbox ;}~3 (21.8, 8.8) \\
{\em j} & 130 & 0.6 & 0.7    & 1.7 & 8.8 & 1.17 &2 (24.2, 18.5){\mbox ;}~3 (27.1, 14.9) \\
{\em k} & 130 & 0.7 & 0.8    & 1.5 & 7.7  & 1.16 &2 (26.5, 18.5){\mbox ;}~3 (29.9, 16.4) \\
{\em l} & 130 & 0.8 & 0.95   & 1.3 & 6.0  & 1.19 &2 (27.0, 14.9){\mbox ;}~3 (31.5, 14.3)\\
\hline
\end{tabular}
\end{table*}

%%% Table 4 %%%

\begin{table*}
\caption{Stability properties of equilibrium models of the galaxy NGC 2985.
The last column gives the pattern speed, $\Omega_p$, and the growth rate Im$\omega$ of the most
unstable $m=2$ and $m=3$ global modes.}

\label{table_4}
\centering
\begin{tabular}{cccccc}
\hline
\hline
Model & $c_r$ &$\sigma_0$ & $M_{disk}$ & $Q_{min}$ & $(\Omega_p$, Im$\omega)$   \\
  & (km s$^{-1}$) & ($10^9 M_{\odot} / kpc^2$) &($10^{10} M_{\odot}$) &  & (\kmskpc)  \\
\hline
{\em a} & 161 & 1.25 & 5.0 & 1.46 & 2(64.6,7.4); 3(98.0,6.0) \\
{\em b} & 161 & 1.5 &  6.2 & 1.19 & 2(79.7,15.9); 3(114.2,12.90 \\
{\em c} & 161 & 1.75 & 7.2 & 1.02 & 2(93.5,24.5); 3(129.2,20.8) \\
{\em d} & 149 & 1.25 & 5.0 & 1.35 & 2(71.3,8.9); 3(110.6,8.2) \\
{\em e} & 149 & 1.5 &  6.2 & 1.10 & 2(87.6,17.8); 3(110.6,8.2) \\
{\em f} & 149 & 1.62 & 6.7 & 1.02 & 2(55.0,22.3); 3(137.0,21.5)  \\
{\em g} & 137 & 1.25 & 5.0 & 1.24 & 2(78.7,10.4); 3(124.7,10.4) \\
{\em h} & 137 & 1.5 &  6.2 & 1.02 & 2(96.5,19.3) 3(146.3,20.8) \\
\hline
\end{tabular}
\end{table*}

%%% Table 5 %%%

\begin{table*}
\caption{Parameters of equilibrium models of the galaxy NGC 3938.
The columns give the vertical velocity dispersion at the disk center,$c_z$,
the ratio of the velocity dispersions $c_z/c_r$, the vertical scale height
of the disk density distribution, $z_0$, the surface density of the disk at the center, $\sigma_0$,
and the minimum value of Q-parameter. The last column gives the
pattern speed $\Omega_p$ and the growth rate Im$\omega$ of the most unstable $m=3$ and
$m=4$ global modes. }

\label{table_5}
\centering
\begin{tabular}{cccccccc}
\hline
\hline
Model & $c_z$ & $c_z/c_r$ & $z_0$ & $\sigma_0$ & $M_{disk}$ & $Q_{min}$ & $(\Omega_p$, Im$\omega)$   \\
 & (km s$^{-1}$) &  & (kpc) & ($10^9 M_{\odot}/ kpc^{2}$) & ($10^{10} M_{\odot}$) &  & (\kmskpc)  \\
\hline
 {\em a}  & 30 & 0.6 & 0.125  & 0.51 & 0.97 & 1.13  & 4(53.9, 9.5) \\
 {\em b}  & 30 & 0.7 & 0.135  & 0.48 & 0.90 & 1.05  & 4(56.5, 10.6) \\
 {\em c}  & 35 & 0.6 & 0.135  & 0.65 & 1.23 & 1.1  & 3(53.7, 13.0); 4(48.1, 13.7) \\
 {\em d}  & 35 & 0.7 & 0.16  & 0.55 & 1.04 & 1.0  & 3(51.8, 8.9); 4(62.1, 13.0) \\
 {\em e}  & 40 & 0.6 & 0.15  & 0.76 & 1.44 & 1.0  & 3(53.6, 15.3); 4(51.6, 15.4) \\
 {\em f}  & 40 & 0.7 & 0.175  & 0.65 & 1.24 & 1.0  & 3(55.6, 14.0); 4(56.4, 16.7) \\
 {\em g}  & 30 & 0.5-0.8 & 0.2  & 0.32 & 0.62 & 1.36-1.92  & stable \\
 {\em h}  & 35 & 0.5-0.8 & 0.2  & 0.44 & 0.84 & 1.16-1.6   & stable \\
 {\em i}  & 40 & 0.5-0.6  &  0.2 & 0.57 & 2.74 & 1.25-1.45 & stable \\
 {\em j}  & 40 & 0.7       & 0.2  & 0.57 & 2.74  & 1.16 & 3(51.3, 8.0); 4(55.6, 9.9)\\
 {\em k}  & 40 & 0.8       & 0.2  & 0.57 & 2.74 & 1.02 & 3(54.9, 11.7); 4(53.3, 13.9)\\
 {\em l}  & 40 & 0.8       & 0.25 & 0.46 & 0.88 & 1.27 & stable \\
\hline
\end{tabular}
\end{table*}

%%% Table 6 %%%

\begin{table*}
\caption{The sequence of equilibrium models
of the galaxy NGC 6503. The columns are: the disk velocity dispersion in radial direction,
$c_r$, the surface density of the disk at the center, $\sigma_0$, the total mass
of the disk, $M_{disk}$, and the minimum value of Q-parameter.
The last column gives the number of arms, the patterm speed, $\Omega_p$, and the growth
rate, Im$\omega$ of the most unstable modes. A few modes grow simultaneously
with close growth rates. }
\label{table_6}
\centering
\begin{tabular}{cccccc}
\hline
\hline
Model & $c_r$ & $\sigma_0$ & $M_{disk}$ & $Q_{min}$ & $(\Omega_p$, Im$\omega)$   \\
  & (km s$^{-1}$) & ($10^9 M_{\odot} /kpc^2$) & ($10^{10} M_{\odot}$) &  & (\kmskpc)   \\
\hline
 {\em a} & 45 & $<$ 0.46  & $<$ 0.39  & $>$ 1.3  & stable \\
 {\em b} & 45 & 0.55  & 0.47  & 1.09  & 3(48.0, 6.8); 4( 44.6, 8.9); 4(54.9, 9.6)\\
 {\em c} & 55 & $<$ 0.6  & $<$ 0.5  & $>$ 1.25  & stable \\
 {\em d} & 55 & 0.62  & 0.52  & 1.19  & 3(47.6, 7.4); 4(41.7, 8.2); 4(62.8, 7.0) \\
 {\em e} & 55 & 0.69  & 0.59  & 1.06  & 3(48.5, 11.8); 4(44.8, 12.5); 4(50.9, 14.0)\\
 {\em f} & 65 & $<$ 0.65  & $<$ 0.55 & $>$ 1.3   & stable \\
 {\em g} & 65 & 0.69  & 0.59 & 1.27  & 3(43.1, 8.1); 4(43.6, 8.3); 4(48.9, 9.0) \\
 {\em h} & 65 & 0.73  & 0.62 & 1.2   & 3 (45.0, 10.2); 4(43.2, 10.3); 4(50.5, 11.0) \\
 {\em i} & 65 & 0.84  & 0.71 & 1.04  & 3(50.5, 17.1); 4(54.8, 18.4); 4(65.2, 18.8) \\
\hline
\end{tabular}
\end{table*}


\begin{thebibliography}{}

\bibitem{}
Adams, F.C., Ruden, S.P., \& Shu, F.H. 1989, ApJ, 347, 959

\bibitem{}
Begeman, K., Broelis, A.H., \& Sanders, R.H. 1991, MNRAS, 249, 523

\bibitem{}
 Bertin, G., Lin, C. C., Lowe, S. A., \& Thurstans, R. P. 1989a,
 ApJ, 338, 78

\bibitem{}
 Bertin, G., Lin, C. C., Lowe, S. A., \& Thurstans, R. P. 1989b,
 ApJ, 338, 104

\bibitem{}
Binney, J., \& Tremaine, S. 1987, Galactic Dynamics,
Princeton University Press, Princeton, New Jersey.

\bibitem{}
Bottema, R. 1992, A\&A, 257, 69

\bibitem{}
Bottema, R. 1989, A\&A, 221, 236

\bibitem{}
Bottema, R. 1988, A\&A, 197, 105

\bibitem{}
Bottema, R. 1997, A\&A, 328, 517

\bibitem{}
Bottema, R., \& Gerritsen, J. P. 1997, MNRAS, 290, 585

\bibitem{}
Dehnen, W, \& Binney, J. 1998, MNRAS, 298, 387

\bibitem{}
Drimmel, R., \& Spergel, D. 2001, ApJ, 556, 181

\bibitem{}
 Carollo, C. M., Stiavelli, M., de Zeeuw, P. T., \& Mack, J.
 1997, AJ, 114, 2366

\bibitem{}
de Grijs, R., Peletier, R. F., \& van der Kruit, P. C.
1997, A\&A, 327, 966

\bibitem{}
de Vaucouleurs, G., \& Caulet, A. 1982, ApJS, 49, 515

\bibitem{}
 Drury, L. 1980, MNRAS, 193, 337

\bibitem{}
 Elmegreen, B. G., \& Elmegreen, D. M. 1990, ApJ, 355, 52

\bibitem{}
 Evans, N. W., Read, J. C. A. 1998, MNRAS 300, 106

\bibitem{}
Freeman, K.C. 1970, ApJ, 160, 811

\bibitem{}
Fuchs, B. 1997, 328, 43

\bibitem{}
 Fuchs, B. 2001a, MNRAS, 325, 1637

\bibitem{}
 Fuchs, B. 2001b, A\&A, 368, 107

\bibitem{}
 Gerssen, J., Kuijken, K., Merrifield, M. R. 1997, MNRAS, 288, 618

\bibitem{}
 Gerssen, J., Kuijken, K., Merrifield, M. R. 2000, MNRAS, 317, 545

\bibitem{}
 Goldreich, P., Lynden-Bell, D. 1965, MNRAS, 130, 125

\bibitem{}
 Grosbol, P. J. 1985, A\&AS, 60, 261

\bibitem{}
Hunter, C. 1979, ApJ, 227, 73

\bibitem{}
 Jim\'enez-Vicente, J., Battaner, E., Rozas, M., et al. 1999, A\&A, 342, 417

\bibitem{}
 Julian, W. H., Toomre, A. 1966, ApJ, 146, 810

\bibitem{}
 Kamphuis, J., \& Briggs, F. 1992, A\&A, 253, 335

\bibitem{}
Karachentsev, I., \& Petit, M. 1990, A\&AS, 86, 1

\bibitem{}
Kent, S.M. 1985, ApJS, 59, 115

\bibitem{}
 Kikuchi, N., Korchagin, V., Miyama, S. M. 1997, ApJ, 478, 446

\bibitem{}
Knapen, J.H., Stedman, S.,  Bramich, D.M.,  Folkes, S.L., \& Bradley, T.R.
2004, AA, 426, 1135

\bibitem{}
 Korchagin, V., Kikuchi, N., Miyama, S. M., et al. 2000, ApJ, 541, 565

\bibitem{}
Korchagin, V.I., Girard, T.M., Borkova, T.V., Dinescu, D.I., \& van Altena, W.F.
2003, AJ, 126, 2896

\bibitem{}
Lacey, C.G. 1984, MNRAS, 208, 687

\bibitem{}
Larsen, S. S., Richtler, T. 1999, A\&A, 345, 59

\bibitem{}
Laughlin, G., Korchagin, V., \& Adams, F. 1998, ApJ, 504, 945

\bibitem{}
 Lin, C. C., Yuan, C., \& Shu, F. H. 1969, ApJ, 155, 721

\bibitem{}
Marochnik, L.S. 1966, AZh, 43, 91

\bibitem{}
Merrifield, M.R. 2001, ASP Conference Series, Vol. 230, 221

\bibitem{}
 Mishurov, Yu. N., Pavlovskaya, E. D., \& Suchkov, A. A.
 1979, AZh, 56, 268

\bibitem{}
Peletier, R.F., Valentijn, E. A., Moorwood, A. F. M., et al.,
1995, A\&A, 300, 1

\bibitem{}
 Persic, M., \& Salucci, P. 1995, ApJS, 99, 501

\bibitem{}
 Peterson, C. J. 1980, AJ, 85, 226

\bibitem{}
 Puerari, I., Dottori, H. A. 1992, A\&AS, 93, 469

\bibitem{}
Puerari, I., Block, D.L., Elmegreen, B.G., Frogel, J.A., \& Eskridge, P.B.
2000, A\&A, 359, 932

\bibitem{}
 Roberts, W. W., Roberts, M. S., \& Shu F. H. 1975, ApJ, 196, 381

\bibitem{}
 Sandage, A., Bedke, J. 1994, The Carnegie Atlas of Galaxies, Volume 1

\bibitem{}
 Sellwood, J. A. 2000, Ap\&SS, 272, 31

\bibitem{}
 Sersic, J. L. 1968, Atlas de Galaxias Australes,
 (Cordoba: Obs. Astron., Univ. Nac. Cordoba)

\bibitem{}
 Shapiro, K. L., Gerssen, J., \& van der Marel R.P. 2003, AJ,

\bibitem{}
Sil'chenko, O.K. 1999, Astron. Lett., 25, 140

%\bibitem{}
% Shu, F. H., Laughlin, G., Lizano, S., Galli, D. 2000, ApJ, 535, 190

\bibitem{}
Sygnet, J. F., Pellat, R., \& Tagger, M. 1987, Phys. Fluids, 30, 1052

\bibitem{}
 Toomre, A. 1964, ApJ, 139, 1217

\bibitem{}
Toomre, A. 1981, The structure and evolution of normal galaxies,
(Cabridge: Cambridge Univ. Press), p 111

\bibitem{}
 Tully, R. B., Verheijen, M. A., Pierce, M. J., et al. 1996, AJ, 112, 2471

\bibitem{}
 van der Kruit, P. C., Freeman, K. C. 1984, ApJ, 278, 81

\bibitem{}
 van der Kruit, P. C., Searle, L. 1982, A\&A, 110, 61

\bibitem{}
 Vauterin, P., \& Dejonghe, H. 1996, A\&A, 313, 465

\bibitem{}
 Vera-Villamizar, N., Dottori, H., Puerari, I., de Carvalho, R. 2001,
 ApJ, 547, 187

\bibitem{}
Villumsen, J.V. 1985, ApJ, 290, 75

\end{thebibliography}
\end{document}